\documentclass[aps,prb,10pt,twocolumn,amsfonts,superscriptaddress,showpacs,floatfix]{revtex4-1}

\usepackage{color,verbatim}
\usepackage{bm}
\usepackage{bbold}
\usepackage{amsmath}
\usepackage{amssymb}
\usepackage{epsfig}

\newcommand{\be}{\begin{equation}}
\newcommand{\ee}{\end{equation}}
\newcommand{\bea}{\begin{eqnarray}}
\newcommand{\eea}{\end{eqnarray}}
\newcommand{\bw}{\begin{widetext}}
\newcommand{\ew}{\end{widetext}}
\newcommand{\ba}{\begin{aligned}}
\newcommand{\ea}{\end{aligned}}

\newcommand{\la}{\langle}
\newcommand{\ra}{\rangle}
\newcommand{\dg}{^\dagger}

\newcommand{\rd}{{\rm d}}

\newcommand{\lm}{\lambda}
\newcommand{\rI}{{\rm int}}
\def\nn{\nonumber\\}

\def\fr#1{(\ref{#1})}

\newcommand{\CoNbO}{${\rm CoNb}_2{\rm O}_6$}

\usepackage[colorlinks,bookmarks=false,citecolor=blue,linkcolor=blue,hyperfootnotes=true]{hyperref}

\begin{document}

\title{Quasiparticle breakdown in the quasi-one-dimensional Ising
  ferromagnet CoNb$_2$O$_6$}
\author{Neil J. Robinson}
\affiliation{The Rudolf Peierls Centre for Theoretical Physics, University of Oxford, Oxford, OX1 3NP, United Kingdom}
\author{Fabian H.L. Essler}
\affiliation{The Rudolf Peierls Centre for Theoretical Physics, University of Oxford, Oxford, OX1 3NP, United Kingdom}
\author{Ivelisse Cabrera}
\affiliation{Clarendon Laboratory, University of Oxford, Parks Road,
Oxford, OX1 3PU, United Kingdom}
\author{Radu Coldea}
\affiliation{Clarendon Laboratory, University of Oxford, Parks Road,
Oxford, OX1 3PU, United Kingdom}

\date{\today}

\begin{abstract}
We present experimental and theoretical evidence that an
interesting quantum many-body effect -- quasi-particle breakdown
-- occurs in the quasi-one-dimensional spin-$\frac12$ Ising-like
ferromagnet \CoNbO\ in its paramagnetic phase at high transverse field 
as a result of explicit breaking of spin inversion symmetry.
We propose a quantum spin Hamiltonian capturing the essential
one-dimensional physics of \CoNbO\ and determine the exchange
parameters of this model by fitting the calculated single particle
dispersion to the one observed experimentally in applied
transverse magnetic fields\cite{CabreraArxiv14}. We present
high-resolution inelastic neutron scattering measurements of the
single particle dispersion which observe ``anomalous broadening''
effects over a narrow energy range at intermediate energies. We
propose that this effect originates from the decay of the one
particle mode into two-particle states. This decay arises from
(i) a finite overlap between the one-particle dispersion and the two-particle 
continuum in a narrow energy-momentum range and (ii) a small misalignment 
of the applied field away from the direction perpendicular to the Ising axis
in the experiments, which allows for non-zero matrix elements for decay by
breaking the $\mathbb{Z}_2$ spin inversion symmetry of the Hamiltonian.
\end{abstract}


\maketitle

\section{Introduction}
Linear spin wave theory and the associated picture of long-lived,
well-defined excitations gives a good description of the static
and dynamic properties of many quantum magnets
\cite{AndersonPR52,KuboPR52,KuboRMP53,DysonPR56a,DysonPR56b,VanKranendonkRMP58,
AkhiezerBook,MattisBook,PauthenetJAP82,ManousakisRMP91}.
Interactions between spin waves can change this picture
substantially and in particular may lead to ``quasi-particle
breakdown''
\cite{ZhitomirskyPRL99,HagiwaraPRL05,VeillettePRB05,SuzukiPRB05,StoneNature06,ZhitomirskyPRB06,MasudaPRL06,KolezhukPRL06,
BibikovPRB07,SyljuasenPRB08,LuscherPRB09,ChernyshevPRB09,MasudaPRB10,FischerNewJPhys10,SyromyatnikovPRB10,StephanovichEPL11,
FischerEPL11,FischerDiss,DorettoPRB12,FuhrmanPRB12,ZhitomirskyRMP13,JoosungPRL13,MourigalPRB13}.
The origin of this effect is that at a given energy and
momentum the single particle mode loses intensity and broadens
significantly as a result of kinematically allowed decay processes
into the multi-particle continua. In contrast to the finite
lifetime of spin excitations induced by scattering with thermal
excitations, quasi-particle breakdown can occur at zero
temperature (see e.g. Ref.~\onlinecite{ZhitomirskyRMP13} for a
recent review). In some cases quasi-particle breakdown is
precluded by a combination of kinematic constraints and the
existence of conservation laws, but can be induced by adding
symmetry breaking terms to the
Hamiltonian\cite{SyljuasenPRB08,LuscherPRB09,ZhitomirskyRMP13}.

While the transverse field Ising chain
(TFIC)\cite{PfeutyAnnPhys70,TFIMBook,SachdevBook} has long been a
key paradigm for quantum phase transitions, an experimental
realization has only been discovered
recently\cite{ColdeaScience10}: the quasi-one-dimensional
Ising ferromagnet \CoNbO\ is formed from
weakly-coupled\cite{CabreraArxiv14} zig-zag chains and exhibits a
phase transition between a spontaneously ordered state
and the quantum paramagnetic phase at an experimentally achievable
critical transverse field of $B_C \approx
5.5$~T\cite{ColdeaScience10}. In the ordered phase weak interchain
couplings give rise to a longitudinal mean field and the resulting
rich spectrum of bound states, predicted 25 years
ago\cite{ZamolodchikovIntJModPhysA89}, has been observed with
inelastic neutron scattering (INS)\cite{ColdeaScience10} and THz
spectroscopy\cite{MorrisPRL14}.

The presence of additional terms in the spin Hamiltonian of
\CoNbO\ beyond the TFIC is under active
investigation\cite{ColdeaScience10,KjallPRB11,CabreraArxiv14}. The
most recent INS study\cite{CabreraArxiv14} has focused on the
high-field paramagnetic phase with the aim of probing the excitations 
in the full Brillouin zone and quantifying the strength of the interchain couplings. 
INS in the paramagnetic phase of the TFIC is expected
to exhibit a sharp high-intensity single particle mode, and low
intensity scattering from the multi-particle
continuum\cite{EsslerReview05,HamerPRB06}. Indeed the INS experiments\cite{CabreraArxiv14}
observed that the excitation spectrum is dominated by a high-intensity single 
particle mode that is sharp over most of the Brillouin zone. 
The parameterization of its dispersion relation indicates that additional terms 
are present in the spin Hamiltonian beyond the leading Ising exchange between
nearest-neighbors along the chain\cite{CabreraArxiv14}. This was also expected based on
a parameterization of the excitation spectrum in zero field
\cite{ColdeaScience10}, numerical studies of the excitation
spectrum in applied field\cite{KjallPRB11}, the value of the
critical field in comparison to the Ising exchange
constant\cite{ColdeaScience10}, and the unusual ``anomalous
broadening'' region seen in INS experiments\cite{CabreraArxiv14}.

In this work we propose a {\em quantitative} one-dimensional
quantum spin Hamiltonian that captures most of
the essential one-dimensional physics of \CoNbO\ in an applied 
transverse field. We determine the parameters of this model by 
fitting the calculated single particle dispersion to the INS data and 
obtain a consistent description of the data at all applied fields tested. 
Having fixed the exchange couplings, we then extend our model in order
to understand the physics behind the ``anomalous broadening'' region
seen in INS scattering -- a narrow energy range at intermediate energies 
across the dispersion bandwidth where the single particle mode is seen to 
broaden and lose intensity. Here we
provide high-resolution INS data for this region, which shows that
the single particle mode has almost vanished. We attribute this to 
quasi-particle breakdown, caused by an overlap between the single 
particle mode and the two-particle continuum and by a small misalignment 
of the applied transverse field, which allows decay processes. 
This interpretation is supported by large scale exact diagonalization studies 
of the quantum spin model with a \emph{single} free parameter, the effective 
misalignment of the magnetic field.

This paper is organized as follows: details of the inelastic
neutron scattering experiments performed on \CoNbO\ are presented
in Sec.~\ref{sec:ExpDetails}. In Sec.~\ref{sec:SpinModel} we
introduce a one-dimensional quantum spin Hamiltonian and detail
the calculation of the single-particle dispersion.
Section~\ref{sec:DSF} explains the fitting procedure used to fix
the exchange parameters of the quantum spin model and studies the
dynamical structure factor of this model using exact
diagonalization. In Sec.~\ref{Sec:Broadening} we present
high-resolution INS data for our study of the ``anomalous
broadening'' region and we present our explanation supported by
exact diagonalization data. Section~\ref{sec:Conclusion} contains
our conclusions and there are two appendices dealing with
technical details underlying our calculations.

\section{Experimental Details}
\label{sec:ExpDetails}
The inelastic neutron scattering measurements of the magnetic
excitations were performed on a 7\,g single crystal of
CoNb$_2$O$_6$ used before [for more details see Ref.
\onlinecite{CabreraArxiv14}] and aligned such that vertical
magnetic fields up to 9\, T were applied along the $b$-axis
(transverse to the Ising axes of all spins). The sample was cooled
to temperatures below 0.06\, K using a dilution refrigerator
insert. The magnetic excitations were probed using the direct
time-of-flight spectrometer LET at the ISIS Facility in the UK,
using neutrons with incident energies of $E_i=4$ and $10$\,meV
with a measured energy resolution [full width at half-maximum (FWHM)]
on the elastic line of 0.051(1) and 0.21(1)\,meV, respectively. LET was operated to
record the time-of-flight data for incident neutron pulses of both
of the above energies simultaneously with typical counting times
of 2 hours for a fixed sample orientation. The higher energy
setting allowed probing the full bandwidth of the magnetic
dispersion along the chain direction $l$ and and the lower energy
setting allowed higher resolution measurements of the low and
intermediate energy ranges to observe clearly the ``anomalous
broadening" effects on the single-particle dispersion. Since we
are mostly concerned here with one-dimensional physics, the
wavevectors are projected along the chain direction~$l$.
\section{One-Dimensional Quantum Model of \CoNbO}
\label{sec:SpinModel}

There now exists extensive experimental evidence
that \CoNbO\ is a quasi-one-dimensional quantum magnet, with 
only small interchain couplings\cite{ColdeaScience10,CabreraArxiv14}. 
With an applied magnetic field along the $b$-axis, \CoNbO\ 
is well described by wealy coupled transverse 
field Ising chains (TFICs)\cite{ColdeaScience10}. 
A microscopic model which attempts to
capture the full one-dimensional (1D) physics of \CoNbO\ must,
however, contain additional interaction
terms\cite{ColdeaScience10,KjallPRB11,CabreraArxiv14}. A natural
first step is to move away from the Ising limit
and consider instead a strongly anisotropic nearest-neighbour XXZ
interaction. The zig-zag crystal structure of the one-dimensional
chains suggests that next-nearest neighbour spin interactions
should also feature in the Hamiltonian, although we expect these
to be weaker due to the longer exchange pathway (Co--O--O--Co
compared to Co--O--Co). Collecting these terms together, we arrive
at a ``minimal one-dimensional spin model'' for \CoNbO: \bea
H &=& H_{TFIC} + H_{XY} + H_{NNN},\label{eq:model}\\
H_{TFIC}  &=& J \sum_\ell S^z_\ell S^z_{\ell+1} + h_x\sum_\ell S^x_\ell,\nn
H_{XY} &=& J\sum_\ell \lm_2 \big( S_\ell^x S_{\ell+1}^x + S_\ell^y S_{\ell+1}^y\big),\nn
H_{NNN} &=& J\sum_\ell \lm_1 S_\ell^z S_{\ell+2}^z + \lm_3 \big( S_\ell^x S_{\ell+2}^x + S_\ell^y S_{\ell+2}^y\big).\nonumber
\eea
Here the $\lambda_i$ are expected to be small, in keeping with the
general arguments presented above and the spin $S=1/2$. The transverse field is
related to the applied magnetic field $B$ by $h_x = g_x\mu_B B$, where
$g_x$ is the g-factor in the $x$ direction. Let us briefly define some
terminology: we will often refer to the Ising easy axis direction $z$
as the ``longitudinal'' direction, whilst the applied field direction
$x$ is the ``transverse'' direction.

A standard approach to calculating the single particle dispersion
of models such as~\fr{eq:model} is linear spin wave theory (see
the data parameterization of Ref.~\onlinecite{CabreraArxiv14}).
This is generally not a reliable approach for one-dimensional
quantum spin models. In the case at hand it permits the
parametrization of the dispersion observed in INS, but requires
different exchange parameters for different values of the
transverse field\cite{CabreraArxiv14}. The origin of this
inconsistency is that higher order terms in the $1/S$ expansion
cannot be neglected. Here we take a different approach to the
problem, based on the self-consistent perturbative treatment of a
fermionic theory\cite{JamesPRB09}. This approach also allows us
to work at finite temperature. 

Following a sequence of transformations (cf. Appendix A of
Ref.~\onlinecite{JamesPRB09}), presented in detail in
Appendix~\ref{App:SelfConBog}, we obtain a fermion theory exactly
equivalent to~\fr{eq:model} where certain parts of the
interactions in $H_{XY}$ and $H_{NNN}$ have been treated exactly.
The Hamiltonian now takes the form 
\bea H &=& \sum_k
E^{\phantom\dagger}_k a^\dagger_k a^{\phantom\dagger}_k +
\frac{J}{L}\sum_{k_i}V_{2}(\mathbf{k})
a^\dagger_{k_1}a^\dagger_{k_2}a_{-k_3}a_{-k_4} \nn &+&
\frac{J}{L}\sum_{k_i}\Big\{ V_{0}(\mathbf{k})
a^\dagger_{k_1}a^\dagger_{k_2}a^\dagger_{k_3}a^\dagger_{k_4} +
\mathrm{H.c.} \Big\}\nn &+& \frac{J}{L}\sum_{k_i}\Big\{
V_{1}(\mathbf{k})
a^\dagger_{k_1}a^\dagger_{k_2}a^\dagger_{k_3}a_{-k_4} +
\mathrm{H.c.} \Big\}\nn &=&H_0+H_{int}, \label{eq:Bog_Ham} \eea
where $H_0$ denotes the quadratic part of $H$. The vertex
functions $V_i({\bf k}) = V_i(k_1,k_2,k_3,k_4)$ are given in
Appendix~\ref{app:Vertex}, $L$ is the system size (number of sites
in the spin chain) and the single-particle dispersion relation is
\bea E_k &=&  \sqrt{\Big[A_k+ \sum_q\Theta_1(k,q) \Big]^2 +
\Big[B_k + \sum_q \Theta_2(k,q)\Big]^2\ }\nn \label{SelfConE} \eea
with $A_k$, $B_k$ and $\Theta_{1,2}$ defined in
Appendix~\ref{App:SelfConBog}. In the $h_x\to\infty$ limit, the
single particle excitations $a\dg_k$ are formed from spin flips in
the completely polarized state $|\leftarrow_x \ldots
\leftarrow_x\ra$; at finite transverse field ($h_x>h_C$) these
become dressed by quantum fluctuations.

The four-fermion interaction terms in the Hamiltonian~\fr{eq:Bog_Ham}
will be treated perturbatively in the following calculation,
consistent with the assumption that $\lambda_i \ll 1$. It should be
emphasized that this perturbative treatment is not equivalent to
simply treating $H_{XY}$ and $H_{NNN}$ directly in perturbation
theory: parts of these interaction terms have been treated exactly
through the self-consistent Bogoliubov transformation performed in
Appendix~\ref{App:SelfConBog}. We now continue by outlining how 
we calculate the single particle dispersion by inverting Dyson's equation.

\subsection{Calculation of the single particle dispersion}
\label{sec:SelfEnergy}

To zeroth order in perturbation theory, the single particle
dispersion is given by Eq.~\fr{SelfConE}. To take into account the
interaction terms present within the Hamiltonian~\fr{eq:Bog_Ham},
we calculate the first order self-energy corrections to the
Green's functions and obtain the modified single-particle
dispersion by resumming an infinite series of diagrams by solving
Dyson's equation. This perturbative calculation is well controlled
provided the thermal energy $k_BT$ is smaller than the single
particle gap $E_{k=0}$; we focus on the behaviour within the
paramagnetic phase and away from the critical point to fulfill
this criterion. We will see that there is good agreement between
the perturbative calculation and the dispersion extracted from
exact diagonalization in this limit. We don't expect our
calculation to predict with any great accuracy the value of the
critical applied field ($B_C \approx 5.5$~T) as the perturbative
expansion becomes uncontrolled in the vicinity of the critical
point.

We begin by discussing the formalism we use for calculating the
modified single particle dispersion and following this we
calculate the first order contributions to the self-energy and
hence the modified single particle dispersion.

\subsubsection{Formalism}

As the Hamiltonian~\fr{eq:Bog_Ham} does not conserve fermion number,
the imaginary time Green's functions take the form of a $2\times2$ matrix
\bea
&&{\bf g}(i\omega_n,k) = - \int_0^\beta d\tau e^{i\omega_n\tau} {\bf g}(\tau,k),\nn
&&{\bf g}(\tau,k) = \left\langle T_\tau \left[ \begin{array}{cc}
a_k(\tau)a^\dagger_k(0)  &  a_k(\tau)a_{-k}(0) \\
a^\dagger_{-k}(\tau) a^\dagger_{k}(0) & a^\dagger_{-k}(\tau)a_{-k}(0) \end{array}
\right] U(\beta) \right\rangle. \nonumber\\
\eea
Here $\tau=it$, $T_\tau$ denotes time-ordering in imaginary time,
$\omega_n$ are Matsubara frequencies,
\be
U(\beta) = T_\tau \exp\left[-\int^\beta_0 d\tau_1 H_{int}\left(\tau_1\right) \right],
\ee
and the expectation value is
\be
\la {\cal O}\ra= \frac{{\rm Tr}[{\cal O}\ e^{-\beta H}]}
{{\rm Tr}[e^{-\beta H}]}\ ,\quad
\beta=1/k_BT.
\label{eq:thermaltrace}
\ee
The noninteracting Green's functions are given by
\bea
{\bf g}_0(i\omega_n,k) &=& \left[
\begin{array}{cc}
G_0(i\omega_n,k) & 0 \\
0 & -G_0(-i\omega_n,-k)
\end{array} \right],\\
G_0(i\omega_n,k) &=& \frac{1}{i\omega_n - E_k }.
\label{g0}
\eea
The full Green's function obeys the Dyson equation
\be
{\bf g}^{-1}(i\omega_n,k) = {\bf g}_0^{-1}(i\omega_n,k) - {\bf \Sigma}(i\omega_n,k), \label{DysonEq}
\ee
where ${\bf \Sigma}$ are the single-particle self-energies. Inverting \fr{DysonEq}
under the assumptions
$({\bf \Sigma})_{21}=({\bf \Sigma})_{12}^*= -({\bf \Sigma})_{12}$ and
$({\bf \Sigma})_{11}= -({\bf \Sigma})_{22}$, which will be verified
at first order in the subsequent calculation, we obtain
\bea
{\bf g}(i\omega_n,k) &=&
\left[ \begin{array}{cc}
 i\omega_n + E_k + ({\bf \Sigma})_{11} &  ({\bf \Sigma})_{21} \\
 ({\bf \Sigma})_{12} & i\omega_n-E_k + ({\bf \Sigma})_{22} \end{array}\right] \nn
 && \times \frac{1}{(i\omega_n)^2-[E_k+({\bf \Sigma})_{11}]^2  - |({\bf \Sigma})_{12}|^2}.
 \label{gf}
\eea
To first order in perturbation theory the self-energy matrix is
frequency independent, and the renormalized single-particle dispersion
can be read off from the position of the pole in the Green's functions
\bea
\varepsilon_k = \sqrt{[E_k + ({\bf \Sigma}(k))_{11}]^2+|({\bf \Sigma}(k))_{12}|^2}\ .
\label{FI_Dispersion}
\eea
At higher orders in perturbation theory the self-energy matrix becomes
frequency dependent and has additional singularities associated with
multi-particle excitations. We now calculate the self-energy matrix
to first order in perturbation theory.

\subsubsection{First order self-energy corrections}
\begin{figure}
\includegraphics[width=0.15\textwidth]{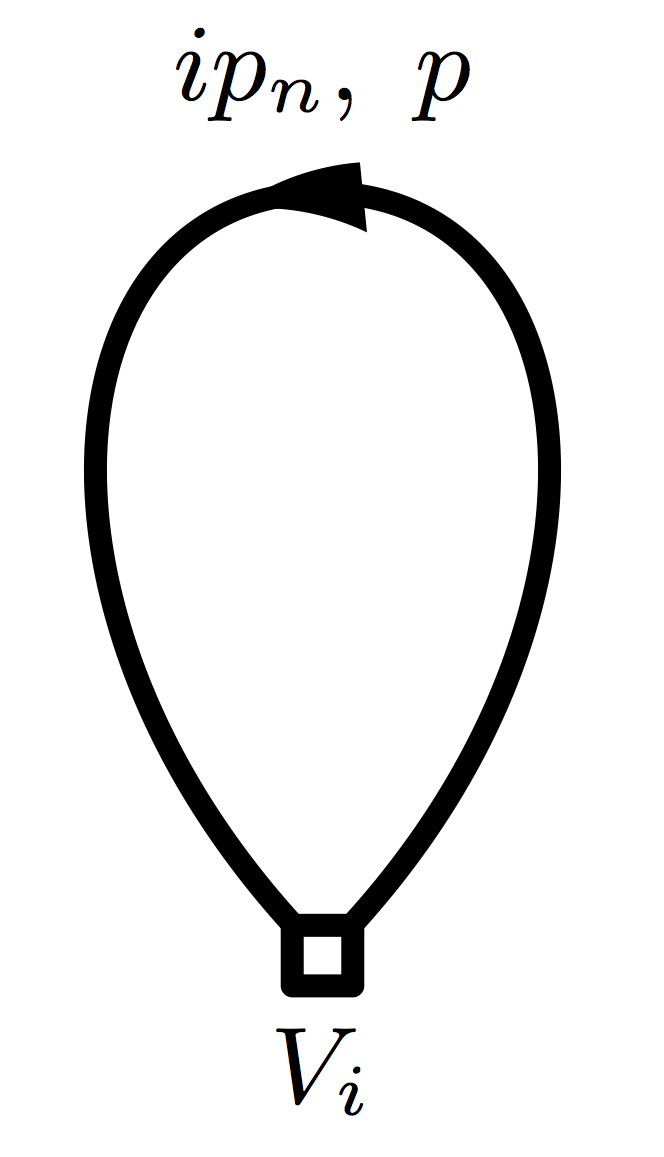}
\caption{The general form of the self-energy diagram at first order. The first-order correction
to the propagators $g^{11}(i\omega_n,k)$ and $g^{22}(i\omega_n,k)$ has $V_i = V_2$,
whilst the anomalous propagators $g^{12}(i\omega_n,k)$ and $g^{21}(i\omega_n,k)$ have
$V_i = V_1$ and $V_i = V_1^*$ respectively. }
    \label{fig:firstorder_gen}
\end{figure}

At first order, the diagrams that contribute to the self-energy are
all of the form presented in Fig.~\ref{fig:firstorder_gen}. We begin
by considering the diagonal matrix elements: the vertex in the self
energy diagram is then given by $V_i = V_2$.
The diagram corresponds to
\bea
({\bf \Sigma}(k))_{11} &=& -\sum_{ip_n,p}\frac{4 J}{\beta L} V_2(k,p,-k,-p) G_0(ip_n,p) e^{ip_n 0^+},\nn
&=& \sum_p 4JV_2(k,p,-p,-k)\frac{n_F(E_p)}{L}, \label{S11fo}
\eea
where $n_F(E_p)=1/(\exp(\beta E_p)+1)$ is the Fermi-Dirac
distribution. The remaining momentum sum in Eq.~\fr{S11fo}
can only be performed numerically, as both the dispersion relation $E_p$
and the vertex function $V_2$ depend upon the Bogoliubov parameter
$\theta_k$, which must be determined numerically from the self
consistency condition~\fr{eq:selfcon}.

We note that from the definition of the self-energy matrix and Eq.~\fr{S11fo} it follows
$({\bf \Sigma}(k))_{11} = -({\bf \Sigma}(k))_{22}$ as the same diagram contributes to
both elements.

The off-diagonal elements of the self-energy matrix are given by the diagram in
Fig.~\ref{fig:firstorder_gen} with $V_i = V_1$ or $V_i = V_1^* = -V_1$. From this,
it follows that $({\bf \Sigma}(k))_{12} = -({\bf \Sigma}(k))_{21}$ and the off-diagonal
self-energy is given by
\be
({\bf \Sigma}(k))_{12} = - 6 \sum_p J V_1(k,-k,p,-p) \frac{n_F(E_p)}{L}. \label{S12fo}
\ee

From Eqs.~\fr{S11fo}--\fr{S12fo} we see that the self-energy is frequency independent
at first order in perturbation theory, hence Eq.~\fr{FI_Dispersion} applies for calculating
the modified single particle dispersion. The elements of the self-energy matrix are
proportional to $J\lm_i n_F(E_p)$; the strongest corrections to the dispersion
occur close to the minima of the dispersion (e.g. in the vicinity of the single particle gap)
or when the system is at high temperatures. The single-particle dispersion with first
order self-energy corrections is given by
\begin{widetext}
\bea
\varepsilon_k
&=& \pm \sqrt{ \Big( E_k + 4J\sum_pV_2(k,p,-p,-k)\frac{n_F(E_p)}{L} \Big)^2
+ \left| 6J\sum_p V_1(k,-k,p,-p)\frac{n_F(E_p)}{L}\right|^2}\ .
\label{fo_disp}
\eea
\end{widetext}
At higher orders in perturbation theory the self-energy matrix becomes
frequency dependent. This introduces additional poles in the Green's
function, corresponding to multi-particle excitations, which can be
determined numerically.

\section{Dynamical Structure Factor}
\label{sec:DSF}

The dynamical structure factor (DSF) $S(\omega,{\bf Q})$ is a
frequency ($\omega$) and momentum (${\bf Q}$) resolved probe of the properties of a magnetic system
\bea
S^{\alpha\beta}(\omega,{\bf Q}) = \frac{1}{L}\int_{-\infty}^{\infty}\rd t\sum_{\ell,\ell'} e^{i{\bf Q}\cdot(\bf{r}_{\ell}-\bf{r}_{\ell'})} e^{i\omega t}
\la S^\alpha_\ell (t) S^\beta_{\ell'} \ra,\nn \label{DSF}
\eea
where $S^\alpha_{\ell}(t) = \exp(iHt)S^\alpha_{\ell}\exp(-iHt)$ is the
time-evolved $\alpha$-component of the spin operator on site ${\bf
  r}_{\ell}$ of the lattice and $\la {\cal O} \ra$ denotes the thermal
trace~\fr{eq:thermaltrace}.  The intensity measured in inelastic
neutron scattering experiments is directly proportional to the
DSF\cite{ZaliznyakBook,ZaliznyakArxiv13}.

The calculation of the DSF for the Hamiltonian \fr{eq:model} is a very
difficult problem. Fortunately we don't require the full solution for
our purposes. The key simplification arises from the fact that both
$S^{zz}$ and $S^{yy}$ are dominated by features due to coherent
single-particle modes, and in fact give the largest contribution to
the measured DSF. These features can be described by a \emph{single-mode
approximation}, which gives a DSF of the form
\bea
S^{\alpha\alpha}(\omega,Q)\Bigg|_{\rm SMA}
&=&A^\alpha(Q)\ \delta(\omega-\epsilon(Q)) \ ,\ \alpha=y,z.
\eea
In the case of the transverse-field Ising chain, the
exact one-particle contributions are known\cite{HamerPRB06}
\bea
A^y(Q)&=&\left[ 1 - \left( \frac{J}{h_x}\right)^2\right]^{1/4}
\epsilon(Q)\ ,\nn
A^z(Q)&=&\left[ 1 - \left(  \frac{J}{h_x}\right)^2\right]^{1/4}\frac{1}{\epsilon(Q)},\nn
\epsilon(Q) &=& \sqrt{h_x^2 - h_x J \cos(Q) + \frac{J^2}{4}}\ .
\eea
We will use that the inelastic neutron scattering data for \CoNbO\ in the
paramagnetic phase exhibits a sharp response along the single particle
dispersion in the $(\omega,Q)$-plane. This allows us (within
experimental resolution) to extract the true single particle
dispersion for excitations in \CoNbO. We then fit the results of our
perturbative calculation~\fr{fo_disp} to the extracted dispersion at a
number of transverse field strengths to consistently fix the exchange
parameters of our model~\fr{eq:model}.

\subsection{Fitting the single particle dispersion to experiment}
\label{subsec:Fit}

\begin{figure*}
\includegraphics[width=\textwidth]{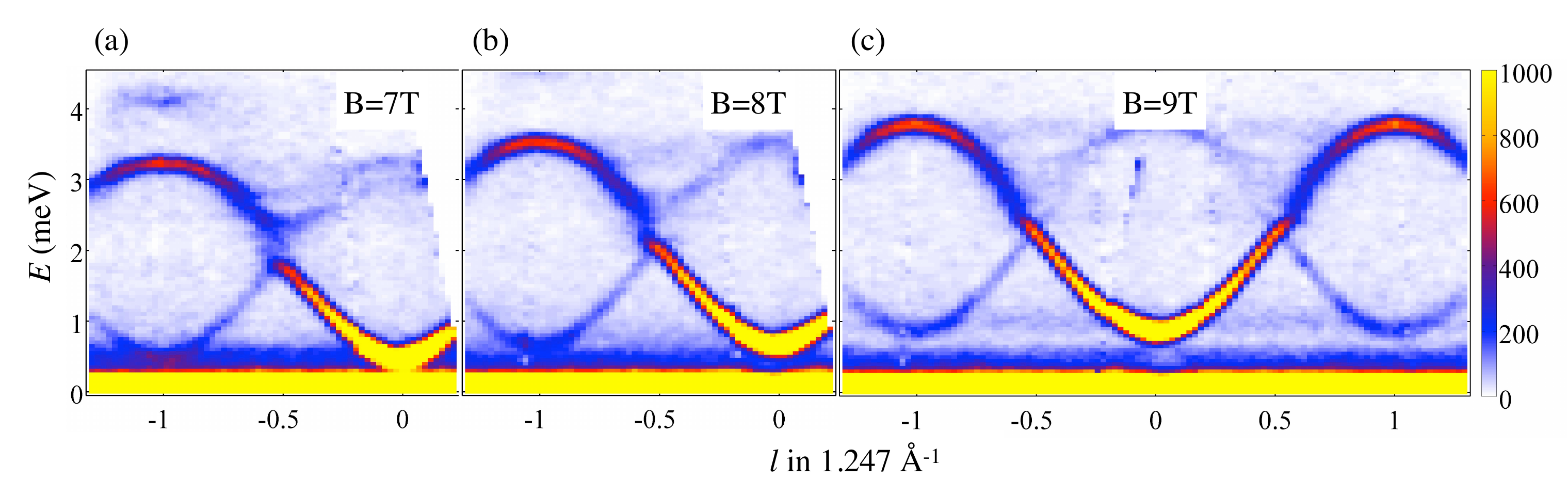}
\caption{(Color online) Inelastic neutron scattering data probing
the dispersion along the chain direction $l$ at (a) $B=7$~T; (b)
$B=8$~T; (c) $B=9$~T. From this data the single particle
dispersion (``Data'' in Fig.~\ref{fig:Fit}) was extracted. Note 
the ``anomalous broadening region'' near $l\approx-0.55$
where the single-particle mode loses weight and significantly
broadens. The incident neutron energy was $E_i = 10$~meV.}
\label{fig:Exp}
\end{figure*}

In Fig.~\ref{fig:Exp} we present inelastic neutron scattering data
for the excitations along the chains for an applied transverse
field of $B=7,8$ and $9$ T. The momentum along the chain direction is
given in reciprocal lattice units of the crystallographic unit
cell along the $c$-direction, i.e. $Q=l 2\pi/c$ where
$2\pi/c=1.247$\AA $^{-1}$. As anticipated in the previous
subsection, the data shows a single sharp quasi-particle
excitation throughout the Brillouin zone (except in the vicinity
of $l\sim-0.55$, which will be discussed later), with additional
weak features due to multi-particle continua. The INS data at
those three fields was parameterized using a 3D dispersion model
(which takes into account also the weak interchain dispersion
normal to the chains as explained in
Ref.~\onlinecite{CabreraArxiv14}), we then extract from this full
parameterization the one-dimensional dispersion along the chain
direction.

We then use a simulated annealing algorithm\cite{NR} to fit the
results of our finite-temperature ($T\approx 50$~mK) perturbative 
calculation~\fr{fo_disp} to the observed one-dimensional single particle 
dispersion for three different values of the applied magnetic field. We run the
simulated annealing algorithm in the $\{\lm_1,\lm_2,\lm_3\}$
parameter space, varying the values of $J$ and $g_x\mu_B$ between
runs and choose a set of parameters which consistently describes
the single particle dispersion across the range of transverse
field strengths. The best fit is obtained for the following set of
parameters: \bea
\begin{aligned}
&J = -2.88~{\rm meV},\quad g_x = 3.21,\\
&\lambda_1 = -0.135,\quad \lambda_2 = 0.205,\quad \lambda_3 = -0.003\ .
\end{aligned}\label{FitParms}
\eea

\begin{figure}
\begin{tabular}{l}
(a) $B=7$T \\
\includegraphics[width=0.45\textwidth]{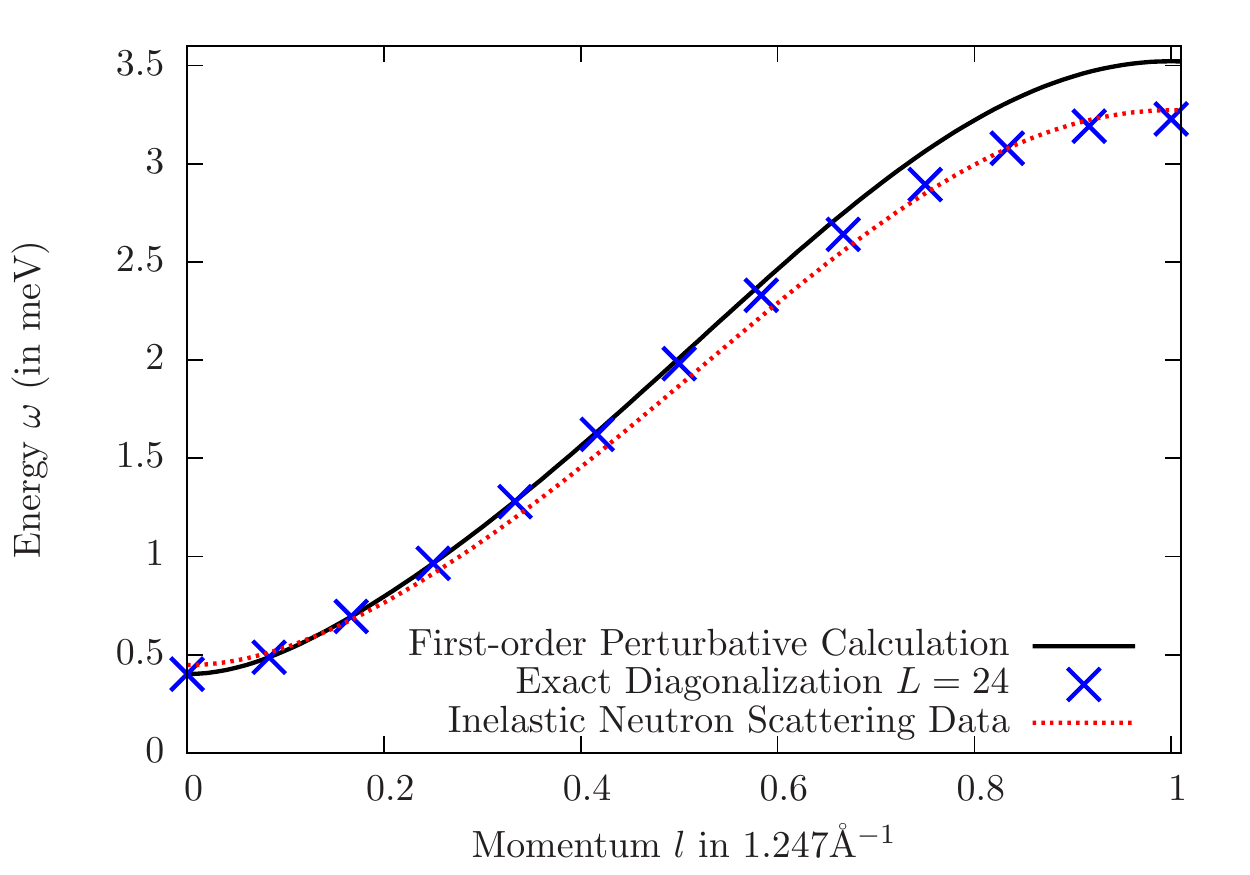} \\
(b) $B=8$T \\
\includegraphics[width=0.45\textwidth]{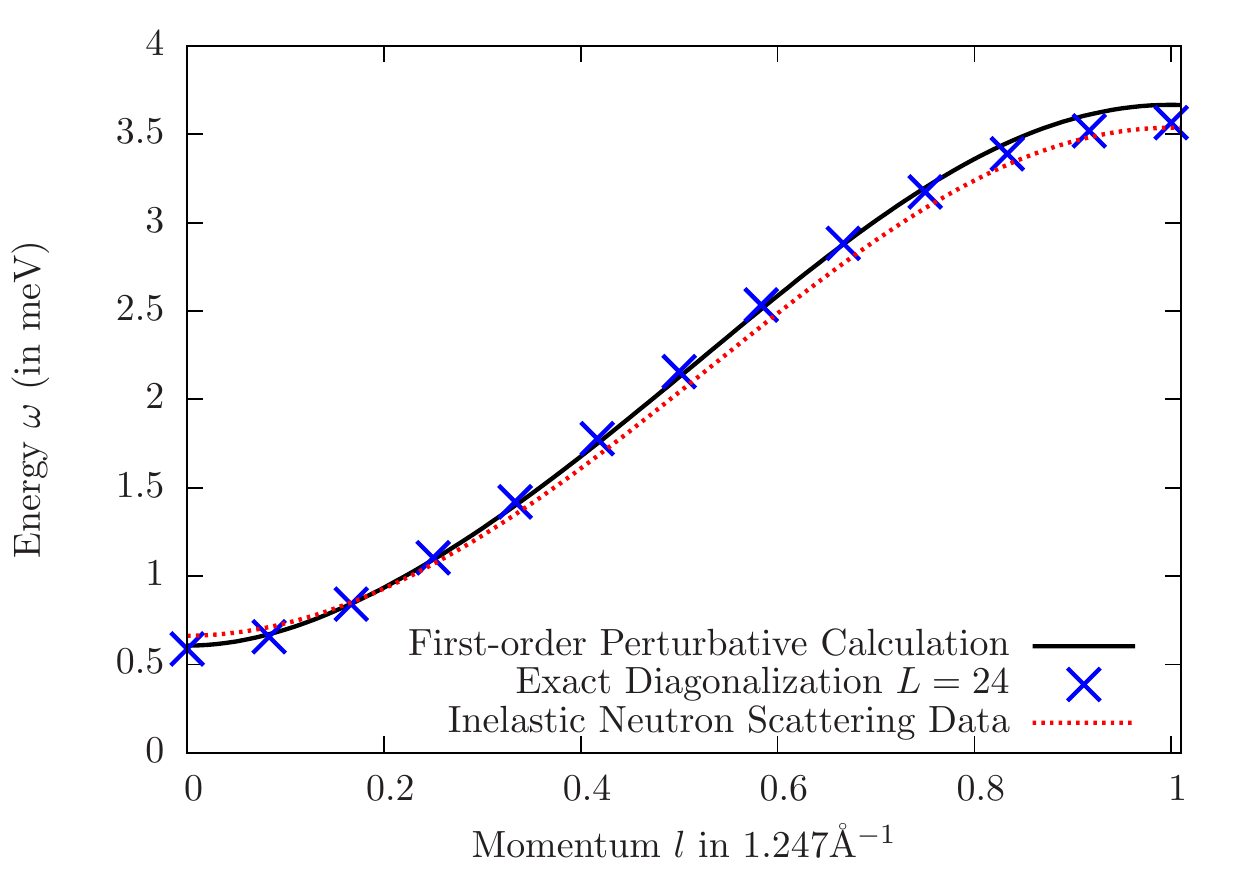} \\
(c) $B=9$T \\
\includegraphics[width=0.45\textwidth]{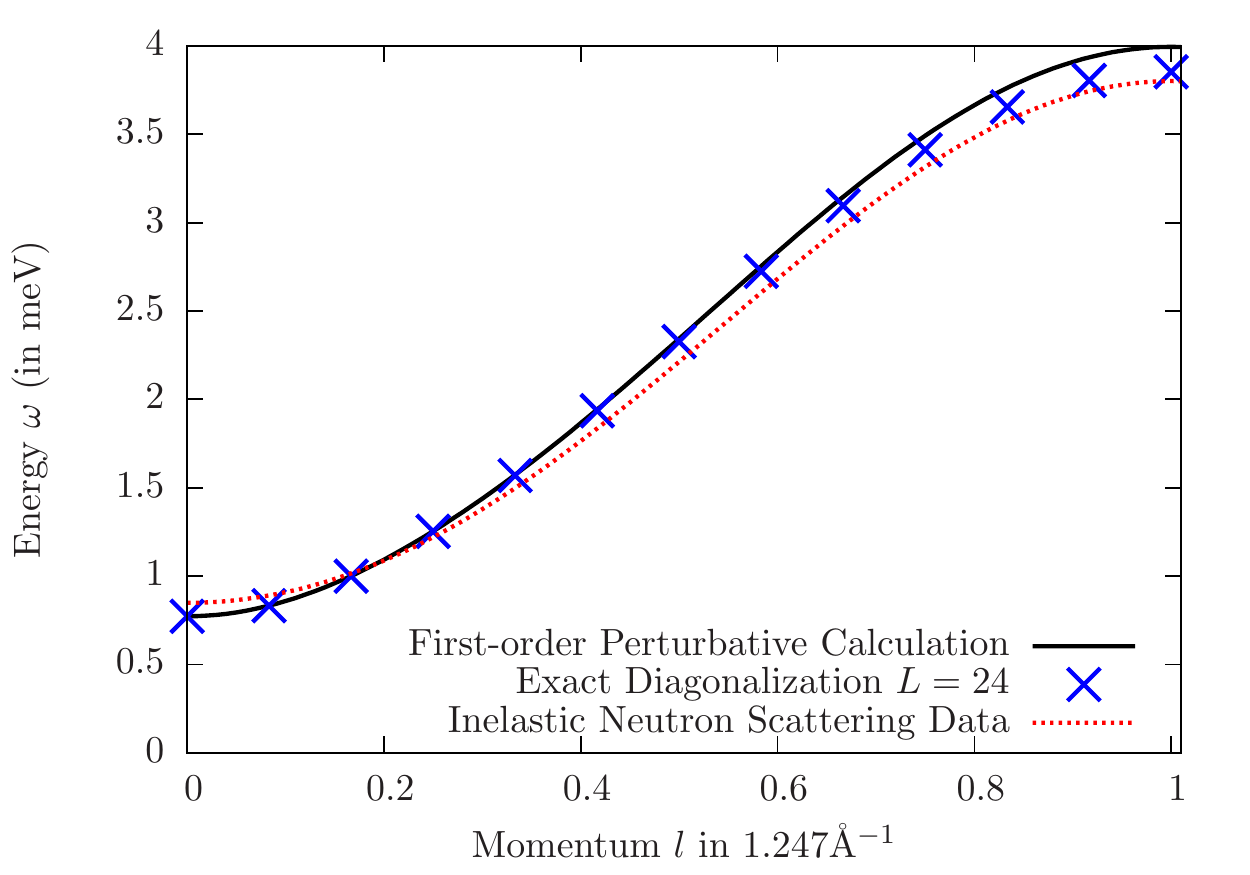}
\end{tabular}
\caption{(Color online) Comparison between the single particle
dispersion calculated by the perturbative calculation at
$T\approx50$~mK (solid line), exact diagonalization of the $L=24$
site system at $T=0$ (blue crosses) and the single particle dispersion
extracted from the inelastic neutron scattering data of
Fig.~\ref{fig:Exp} (dotted line). We see that the perturbative
calculation over estimates the single particle dispersion at
$l\approx 1$ (especially for $B=7$~T), nevertheless exact
diagonalization results match the experimental data very well.}
\label{fig:Fit}
\end{figure}

Comparisons between the calculated single particle dispersion
(solid line), exact diagonalization results for the
Hamiltonian~\fr{eq:model} with the above parameters and the
extracted parameterization of the dispersion from inelastic
neutron scattering data (Fig.~\ref{fig:Exp}) (dotted line) are
shown in Figs.~\ref{fig:Fit}(a)-(c). We see that the perturbative
calculation overestimates the single particle dispersion at
$l\approx 1$ for $B=7$~T, but the exact diagonalization results are
in excellent agreement with the experimental data for all fields.
The perturbative calculation allows us to estimate the critical
transverse field: the parameter set~\fr{FitParms} leads to a
one-dimensional critical field strength of $h_C = 0.915$~meV
($B^{1D}_C \sim 4.92$~T), i.e. the field where the one-dimensional
chains would have been critical in the absence of inter-chain
couplings. We stress that our perturbative calculation is not
controlled in the vicinity of the critical point, but this value
broadly agrees with the experimental estimate of the 1D critical
field\cite{ColdeaScience10}. The perturbative result for the critical
field is also in excellent agreement with the field $h_C = 0.908$~meV 
at which the extrapolated ($L=\infty$) single-particle gap vanishes in 
exact diagonalization studies of the Hamiltonian~\fr{eq:model} with
parameters~\fr{FitParms}.

In the following, we will use the parameter set \fr{FitParms} to
carry out exact diagonalization studies of the DSF. Comparing
these results to the INS data will lend further support to our
claim that the model~\fr{eq:model}, \fr{FitParms} gives a good
description of the one-dimensional physics of \CoNbO.

\subsection{Exact diagonalization: Eigenvalue Spectrum}
\label{sec:DSF_ED}

\begin{figure}
\begin{tabular}{l}
(a) $B=7$T\\
\includegraphics[width=0.45\textwidth]{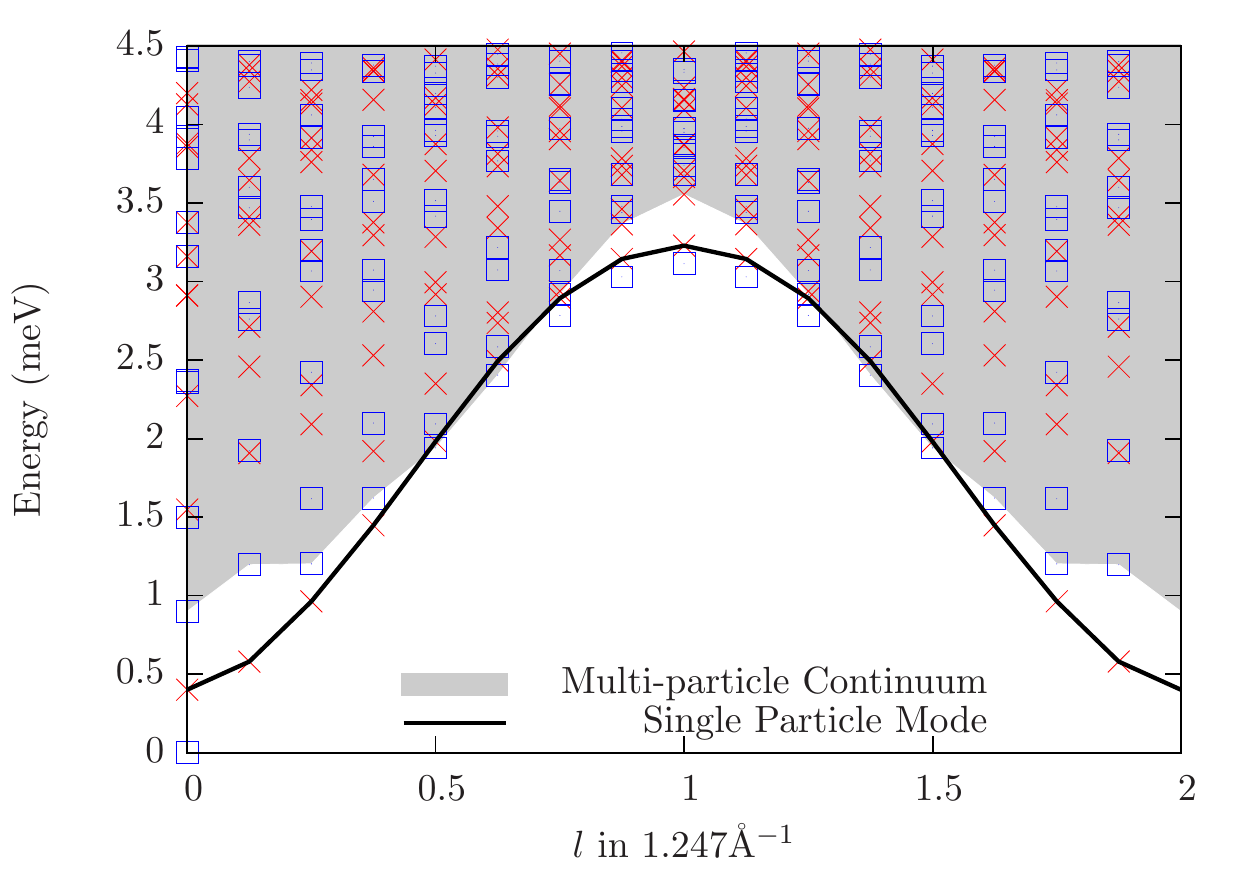} \\
(b) $B=8$T\\
\includegraphics[width=0.45\textwidth]{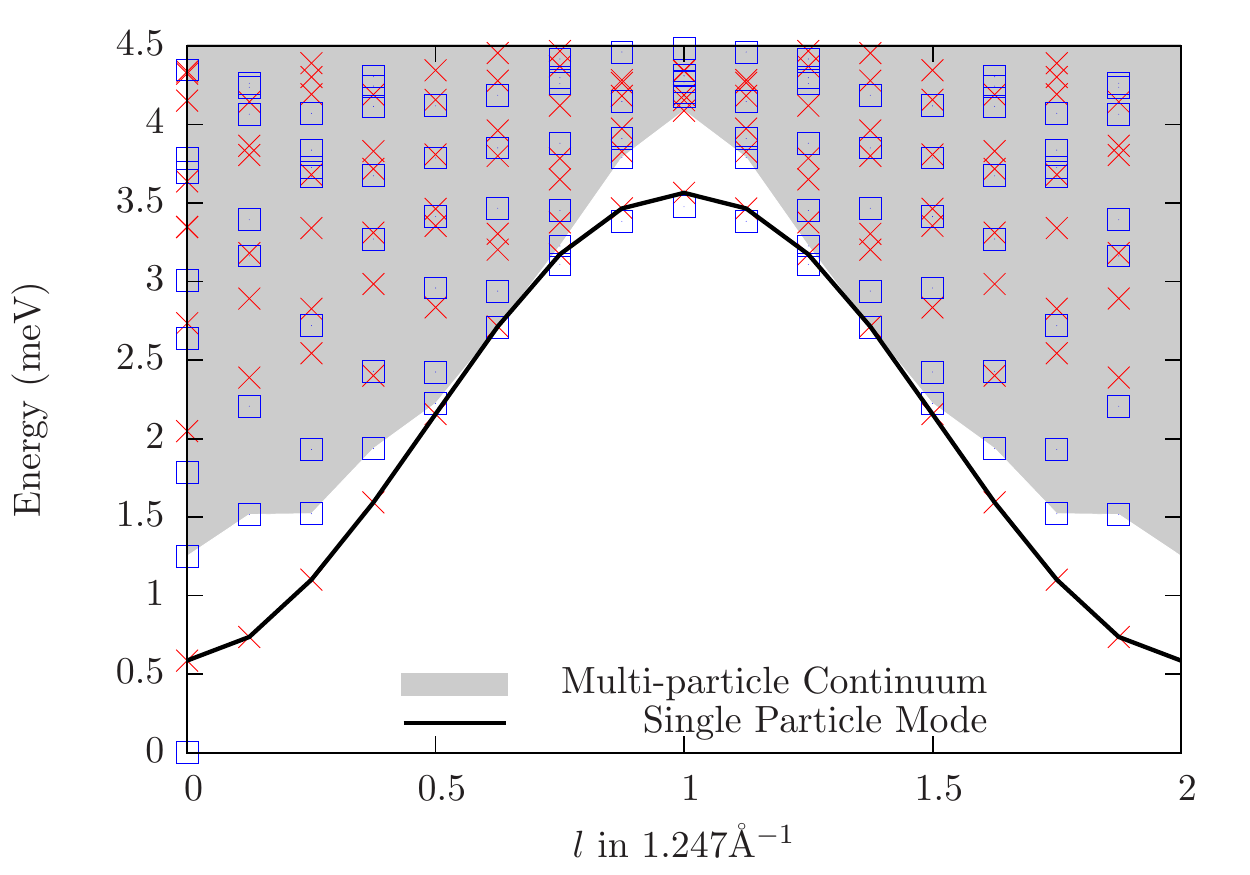} \\
(c) $B=9$T\\
\includegraphics[width=0.45\textwidth]{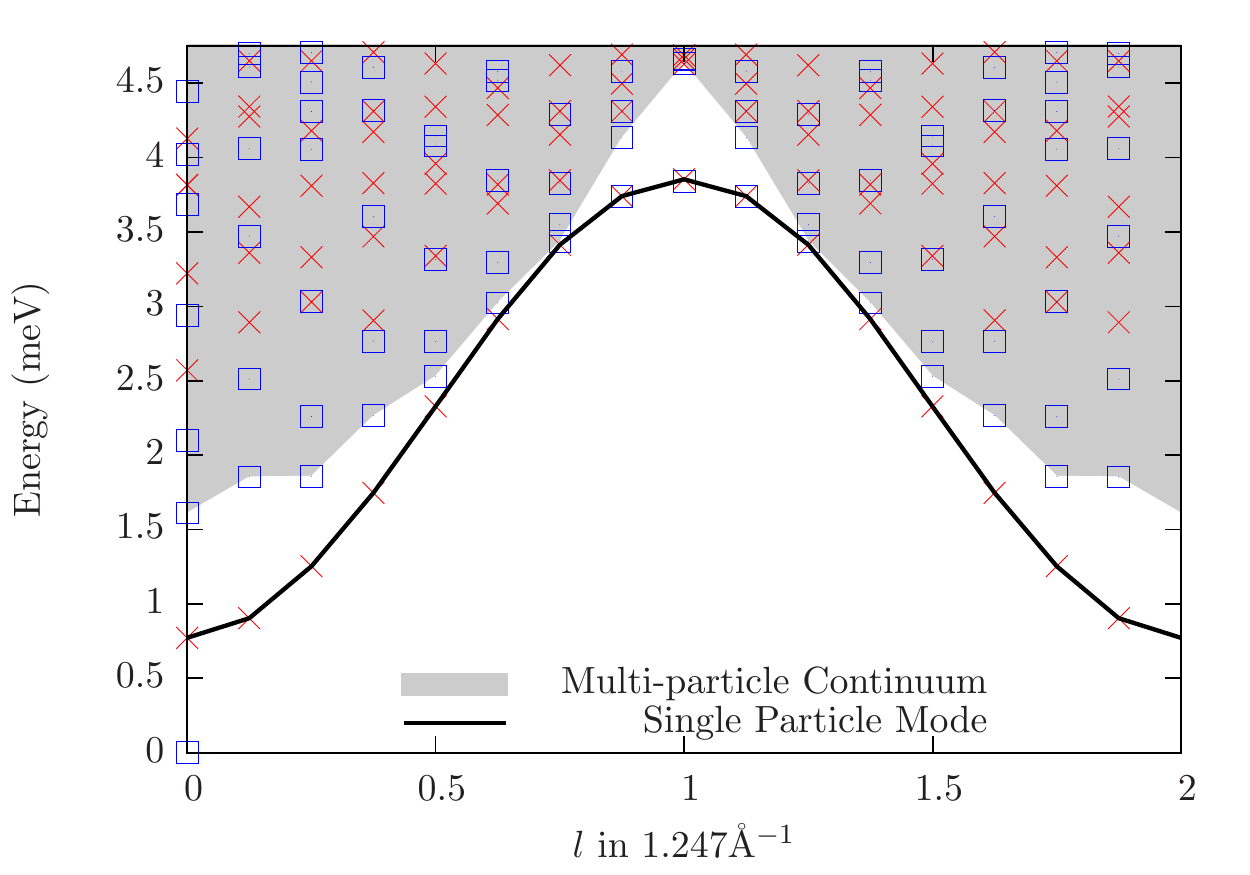}
\end{tabular}
\caption{(Color online) We present the spectrum of the
Hamiltonian~\fr{eq:model} with parameters~\fr{FitParms} obtained
by exact diagonalization of the $L=16$ chain at (a) $B=7$~T; (b)
$B=8$~T and (c) $B=9$~T. The parity under spin inversion
$S^z\to-S^z$ of each state is labelled by crosses (odd) and
squares (even). In particular we highlight the single particle
mode (SPM) (solid line) and the multi-particle continuum (shaded
region), showing that the SPM is close to or overlapping with the
continuum for $l\approx0.5-0.6$ in all three cases. There is a
two-particle bound mode (blue squares below the continuum
boundary) near the ferromagnetic zone boundary ($l=1$) with a
similar energy to the SPM.} \label{fig:Z2_Spectrum}
\end{figure}

We start by considering the spectrum of the spin
model~\fr{eq:model}, obtained by fully diagonalizing the
Hamiltonian. This will be useful for our discussions of the DSF,
particularly in describing the unusual broadening region (see
Sec.~\ref{Sec:Broadening}). Figures~\ref{fig:Z2_Spectrum}(a)--(c)
present the spectrum of the Hamiltonian for $B=7,8,9$~T, where we
have specified the symmetry of each state under spin inversion
$S^z_i \to -S^z_i$. The single particle mode is shown as a solid
line, while the extent of the multi-particle continua is indicated
by the grey shaded region. In all three cases we see that the
single particle mode grazes the two-particle continuum in the
region $l=0.5-0.7$, with the three-particle continuum also close
by at lower fields (within $\sim0.25$~meV at $B=7$~T). This
overlapping of the single particle mode with the multi-particle
continuum is a result of physics beyond the transverse field Ising
chain, for which this cannot occur in the paramagnetic phase due to
kinematic constraints enforcing $E_{k} + E_{q-k} > E_k$ for all
$k,q$.

\subsection{Lanczos diagonalization: The DSF}
\label{Sec:LancDSF}

Having examined the spectrum of the Hamiltonian, we next turn our
attention to the DSF. To study the DSF, we move away from full
diagonalization of the Hamiltonian and use Lanczos based
techniques to iteratively diagonalize the Hamiltonian, allowing us
to work on much larger system sizes (up to $L=28$, where each
momentum block of the Hamiltonian has dimension $\approx 2^{28}/28
= 9.6\times10^6$). This significantly increases our momentum and
frequency resolution, which will be useful in particular for
examining the anomalous broadening region. We use that the
diagonal components of the structure factor~\fr{DSF} can be
written as \be S^{\alpha\alpha}(\omega,Q) =  \frac{1}{\pi}
\lim_{\eta \to 0}\ {\rm Im}\ \la S^{\alpha}_{Q}| \frac{1}{\omega +
i\eta + E_0 - H} |S^{\alpha}_Q\ra, \nonumber \ee where
$S^{\alpha}_Q$ is the Fourier transform of the spin operator
$S^{\alpha}_l$, $|S^\alpha_Q\ra$ is the ground state with the
Fourier transformed spin operator applied to it and $E_0$ is the
ground state energy. In our numerics we take $\eta = 0.01J$, which
broadens the delta-functions peaks of the DSF by a Lorentzian.

Our procedure for calculating the diagonal components ($\alpha =
x,y,z$) of the DSF is as follows: (i) we begin by using a Lanczos
procedure to find the ground state; (ii) we construct the state
obtained by acting on the ground state with the Fourier
transformed spin operator; (iii) we perform an additional Lanczos
procedure with the constructed state as the initial state and then
calculate the DSF using the continued fraction
representation\cite{GaglianoPRL87,DagottoRMP94}.

\begin{figure}
\begin{tabular}{l}
(a)  \\
\includegraphics[width = 0.48\textwidth]{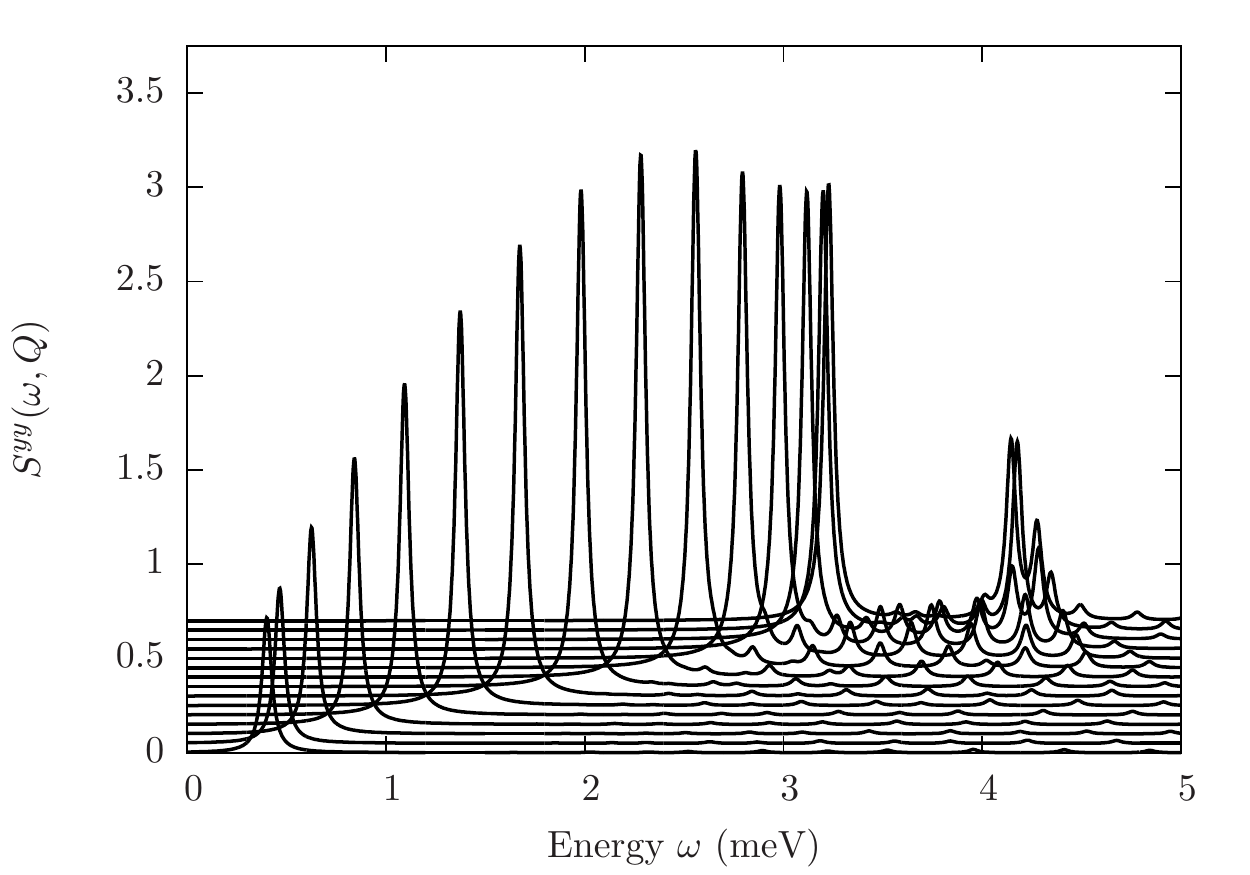} \\
(b) \\
\includegraphics[width = 0.48\textwidth]{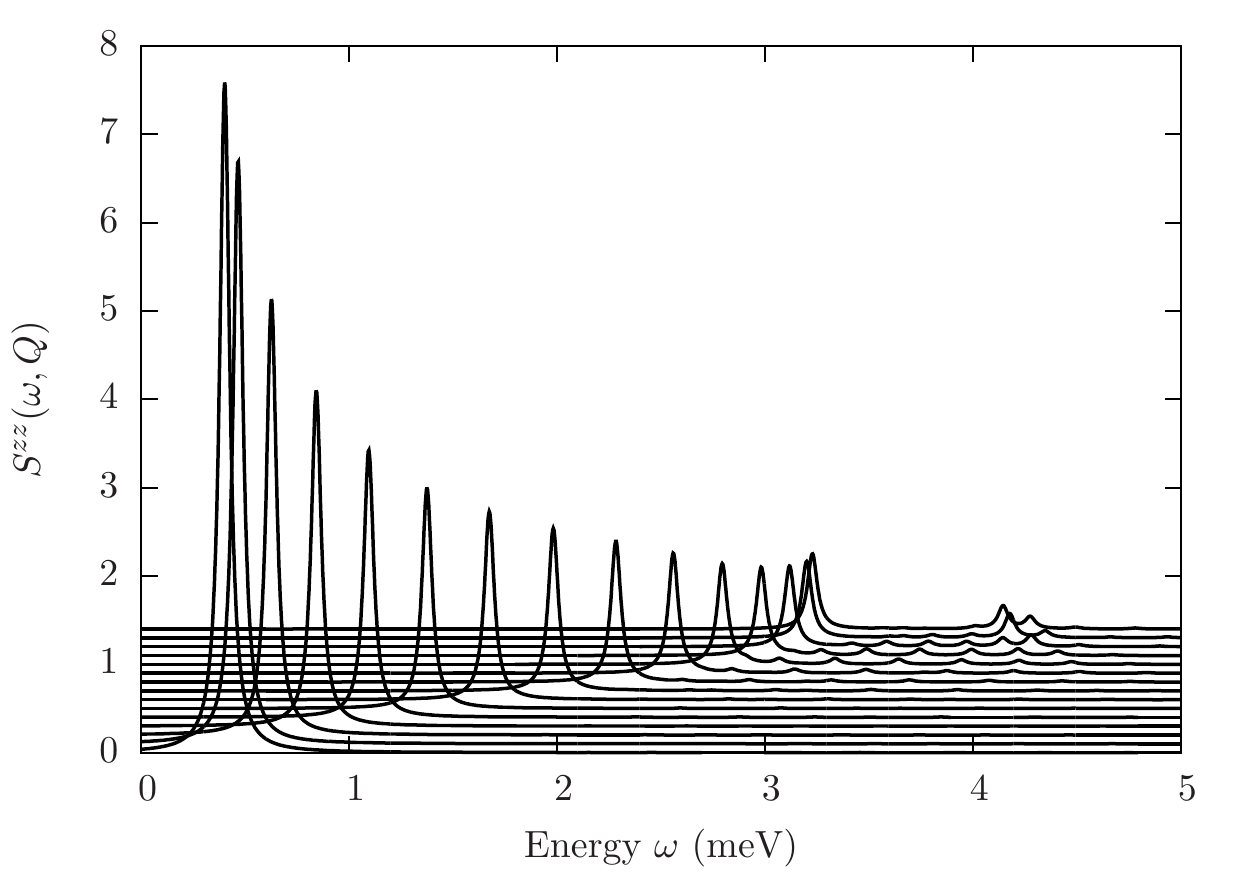}
\end{tabular}
\caption{Constant wave vector cuts ($l=0\to1$ in steps of $\delta l = 1/14$,
vertically displaced for clarity) of the dynamic structure factor 
$S^{\alpha\alpha}(\omega,Q=l 2\pi/c)$ for (a) $\alpha = y$ and (b) $\alpha=z$
at $B=7$~T on the $L=28$ chain with Hamiltonian~\fr{eq:model} and exchange
parameter~\fr{FitParms}. We have used $120$ Lanczos iterations in
the continued fraction and broadening parameter $\eta = 0.01J$. }
\label{fig:DSF}
\end{figure}

Following this procedure we find the DSF of the
Hamiltonian~\fr{eq:model} with exchange parameters~\fr{FitParms} for
$B=7,8,9$~T. We present the data for $B=7$~T in Fig.~\ref{fig:DSF},
where we have focussed on the $\alpha = y,z$ components of the DSF as
these carry most of the spectral weight. The DSF is dominated
by a single sharp mode across the Brillouin zone, with the multi-particle continua 
having non-negligible weight at $l\approx1$ and
$E\approx 4$~meV. This should be compared to the INS data presented in
Fig.~\ref{fig:Exp}, where a similar feature is observed. As seen in
experiment, with increasing applied transverse field $B$ the multi-particle 
feature moves to higher energies and becomes less
intense. The single particle mode also moves up in energy
with applied transverse field, as depicted in Figs.~\ref{fig:Fit}.

We see that whilst both the general features and the quantitative
behaviour with transverse field of the DSF are captured by the
minimal one-dimensional spin model~\fr{eq:model}, we \emph{do not} see the
anomalous broadening region observed in
experiments\cite{CabreraArxiv14}, see Fig.~\ref{fig:Exp}. In the
next section we present high-resolution INS data for this
phenomenon and propose a likely explanation of its origin.

\section{Anomalous broadening and quasi-particle breakdown}
\label{Sec:Broadening}

\subsection{High resolution inelastic neutron scattering: Broadening region}

\begin{figure*}
\includegraphics[width=\textwidth]{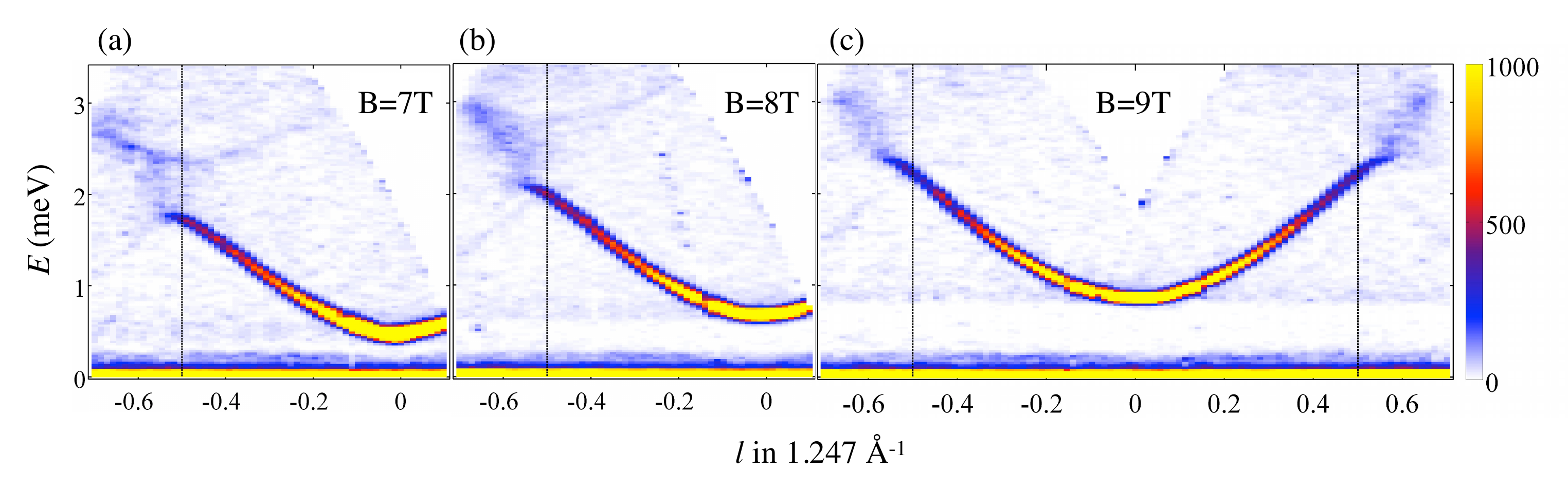}
\caption{(Color online) High-resolution inelastic neutron
scattering data for the single particle dispersion with momentum
oriented along the chain. Note that the ``anomalous broadening"
region where the sharp mode loses weight and disappears is located
distinctly away from the crystallographic zone boundary positions
$l=\pm0.5$ emphasized by vertical dotted lines. The data was
obtained for neutrons with an incident energy of $E_i = 4$~meV and
a corresponding resolution on the elastic line of $\Delta E =
0.051(1)$ meV. Data is shown for three applied transverse field
strengths: (a) $B=7$~T; (b) $B=8$~T and (c) $B=9$~T.}
\label{fig:AnomBroad}
\end{figure*}

A surprising feature of the INS data shown in Fig.~\ref{fig:Exp},
is that close to $l\approx0.5$ the single particle mode appears to
broaden and lose a significant amount of weight.
Figure~\ref{fig:AnomBroad} presents high-resolution INS data (with
resolution on the elastic line of $\Delta E = 0.051(1)$~meV (FWHM))
focussed on this particular feature. The broadening and reduction
in weight is so extreme, that at $B=7$~T a gap appears to have
opened in the single particle mode; a careful analysis of the data
shows that this feature does \emph{not} occur at $l=-0.5$ but at
wavevectors distinctly away from it (most clearly seen in Fig.\
\ref{fig:AnomBroad}, the ``anomalous broadening" occurs
away from the crystallographic zone boundary points $l=\pm0.5$
indicated by vertical dotted lines). Hence it cannot be attributed
to a zone boundary gap due to a doubling of the unit cell, such as
seen in dimerization transitions (e.g. a Peierls transition\cite{PeierlsMoreSuprises}).

\begin{figure*}
\includegraphics[width=0.8\textwidth,trim = 5mm 5mm 0mm 2mm,clip=true]{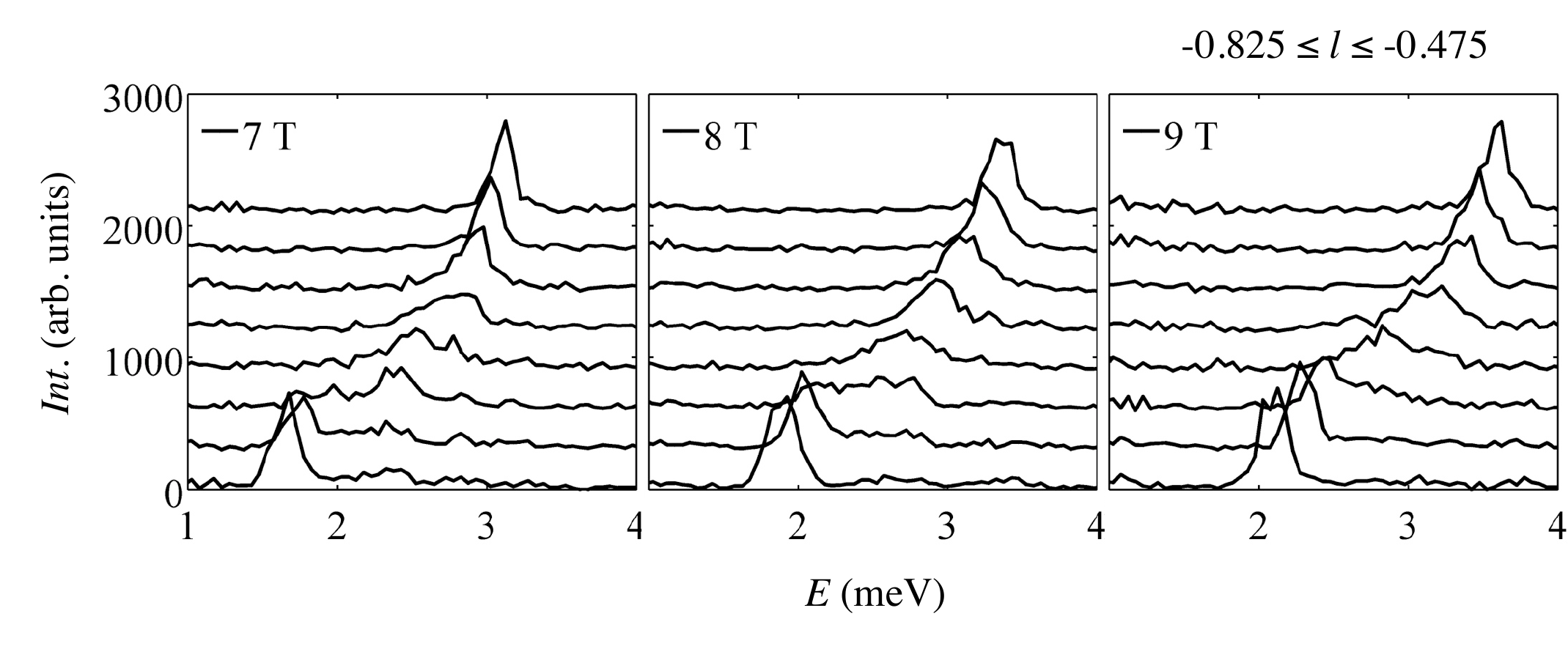}
\caption{Scans through the neutron scattering data in Fig.~\ref{fig:Exp} highlighting the anomalous
broadening of the single particle mode in the region near $l\approx-0.55$. 
Traces (offset vertically and excluding error bars for clarity) correspond to the intensity as a function of 
energy in scans at fixed momentum l in the range $l =-0.475$ (lowest trace) to $l = -0.825$ (highest trace) 
in steps of 0.05 (each with an integration range of $\delta l = \pm0.025$ around the nominal $l$-value) 
at  (left) $B = 7$~T, (center) $B = 8$~T, and (right) $B = 9$~T. 
Note the broadening of the peaks for energies $E(7~{\rm T}) \approx 2.0 - 2.75$~meV, 
$E(8~{\rm T}) \approx 2.25 - 3.0$~meV, and  $E(9~{\rm T}) \approx 2.5 - 3.25$~meV.}
\label{fig:cuts}
\end{figure*}

The change in the magnetic scattering intensity as a function of energy and momentum
is shown in a series of constant-momentum cuts in Fig.~\ref{fig:cuts}, where we focus on the 
region of broadening $-0.825 < l < -0.475$. The largest broadening and reduction of weight 
occurs when $B = 7$~T in the energy range $E(7~{\rm T}) \approx 2.0 - 2.75$~meV. 
At higher magnetic fields these features become less pronounced but are still
clearly visible, with broadening observed for energies $E(8~{\rm T}) \approx 2.25 - 3.0$~meV, 
and  $E(9~{\rm T}) \approx 2.5 - 3.25$~meV.

\subsection{Broadening of the single particle mode at intermediate energies}

In the remainder of this paper, we focus on explaining the
``anomalous broadening'' region in the INS data. The spin model
introduced in Sec.~\ref{sec:SpinModel} and the fit parameters of
Sec.~\ref{subsec:Fit} serve as a starting point for exact
diagonalization studies. As we have seen in the previous section,
the DSF for the Hamiltonian~\fr{eq:model} is dominated by a single
dispersive mode that is sharp across the whole Brillouin zone and
so does not capture the physics of the broadening of the single
particle mode see in experiments. To go beyond this, we take
inspiration from the data presented in Figs.~\ref{fig:Z2_Spectrum}(a)--(c), 
which show that the single particle mode and the multi-particle continuum 
overlap in the same region as the anomalous broadening is observed 
in the INS data. We also observe that the multi-particle excitations which 
are in the vicinity of the single particle mode are even under spin inversion
symmetry, whilst the single particle mode is itself odd. As a result, 
transitions between the single particle mode and close by multi-particle
excitations are forbidden in the Hamiltonian~\fr{eq:model}. 
Importantly, Figs.~\ref{fig:Z2_Spectrum}(a)--(c) also show
that the multi-particle excitations in the vicinity of the single particle 
dispersion are even under spin inversion $S^z\to-S^z$, whilst the 
single particle mode is odd and so mixing of the
two types of excitation is disallowed by the $\mathbb{Z}_2$ symmetry of
the Hamiltonian. 
With this in mind, we add an additional term to the
Hamiltonian~\fr{eq:model} which breaks the $\mathbb{Z}_2$ spin
inversion symmetry $S^z_i \to -S^z_i$ of the model: A natural 
candidate for such a term is a small \emph{longitudinal field}  
$h_z = g_z \mu_B B_z$ which would arise in the experimental 
setting due to not having perfect alignment of the crystal with respect 
to the transverse field.\footnote{One may think that off-diagonal elements 
of the $g$-tensor might have the same effect. However, as a result of the 
local symmetry point group at the Co$^{2+}$ site (two-fold rotation axis around $b$), 
the $b$-axis is a principle axis of the $g$-tensor so an external magnetic field applied
strictly along the $b$-axis does not induce a longitudinal field component.}
Thus we consider the Hamiltonian modified by \bea H \to H + h_z
\sum_l S^z_l\ . \label{eq:pert} \eea For the inelastic neutron
scattering data presented in
Figs.~\ref{fig:Exp},~\ref{fig:AnomBroad} and~\ref{fig:cuts}, it is estimated that
the crystal was aligned such that the magnetic field was
perpendicular to the Ising axis to within an accuracy of
$\sim1^\circ$.

It is worth noting that transitions between the 1 and 3 particle
states can occur without the breaking of $S^z$ spin inversion
symmetry. However, as can be seen in
Figs~\ref{fig:Z2_Spectrum}(a)--(c), the three particle states are
kinematically well separated from the single particle mode (no
overlap), and decay $1\rightarrow 3$ can therefore not account for
the anomalous broadening.

We also wish to highlight the fact that the overlap of the
one-particle mode with the multi-particle continua does not occur
within the paramagnetic phase of the transverse field Ising chain
($\lambda_{1} = \lambda_2 = \lambda_3 = 0$):
The overlap occurs in the present case due to the additional
exchange interactions present in the Hamiltonian~\fr{eq:model}
which modify the dispersion shape such that an overlap of one and
two-particle states exists for a finite field range above the
critical field.

Let us now briefly summarize the requirements for the broadening
of the single particle mode:
\begin{enumerate}
\item The single particle mode and the multi-particle continuum must
  overlap (see Figs.~\ref{fig:Z2_Spectrum}(a)--(c)).
\item Matrix elements must exist between the single particle mode and
the overlapping states within the multi-particle continua. If these
states are two-particle states, the $S^z$ spin inversion symmetry must
be broken to allow transitions.
\item The decay rate of the single particle mode must be sufficiently
large for the broadening to become apparent.
\end{enumerate}

\subsection{Lanczos Diagonalization (up to $L=28$)}
\label{AnomBroadLanczos}
We now turn to exact diagonalization results for the DSF in the
presence of a small longitudinal field. As the broadening effect
that we are looking for is seen in a certain area of the Brillouin
zone, we use Lanczos diagonalization (and associated continued
fraction techniques\cite{GaglianoPRL87,DagottoRMP94})  to extend
the momentum resolution of our calculations (for full
diagonalization we are limited to $L\sim 18$ sites). We focus on
the diagonal components of the DSF $S^{\alpha\alpha}(\omega,Q)$
with $\alpha=y,z$ as these carry most of the intensity. To allow
us to compare the regions of anomalous broadening for different
strength of the transverse field, we work with a fixed ``crystal
misalignment'' of $\theta \sim 1.5^\circ$, and we use $g_z = 5.9$
(we estimate from Ref.~\onlinecite{KunimotoJPSJ99} that $g_z
\approx 5.6-6.2$).

\begin{figure*}
\begin{tabular}{ll}
(a) & (b) \\
\includegraphics[width=0.48\textwidth]{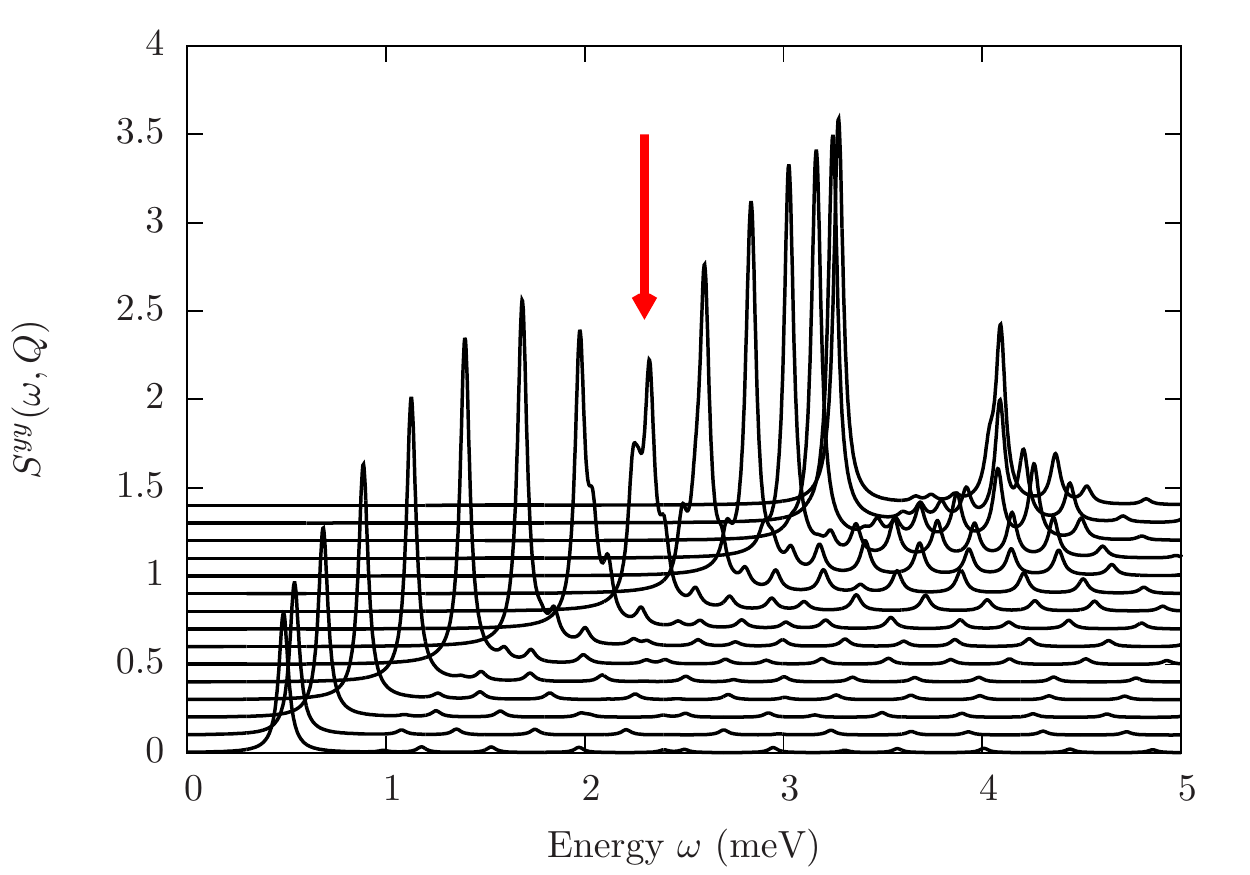} &
\includegraphics[width=0.48\textwidth]{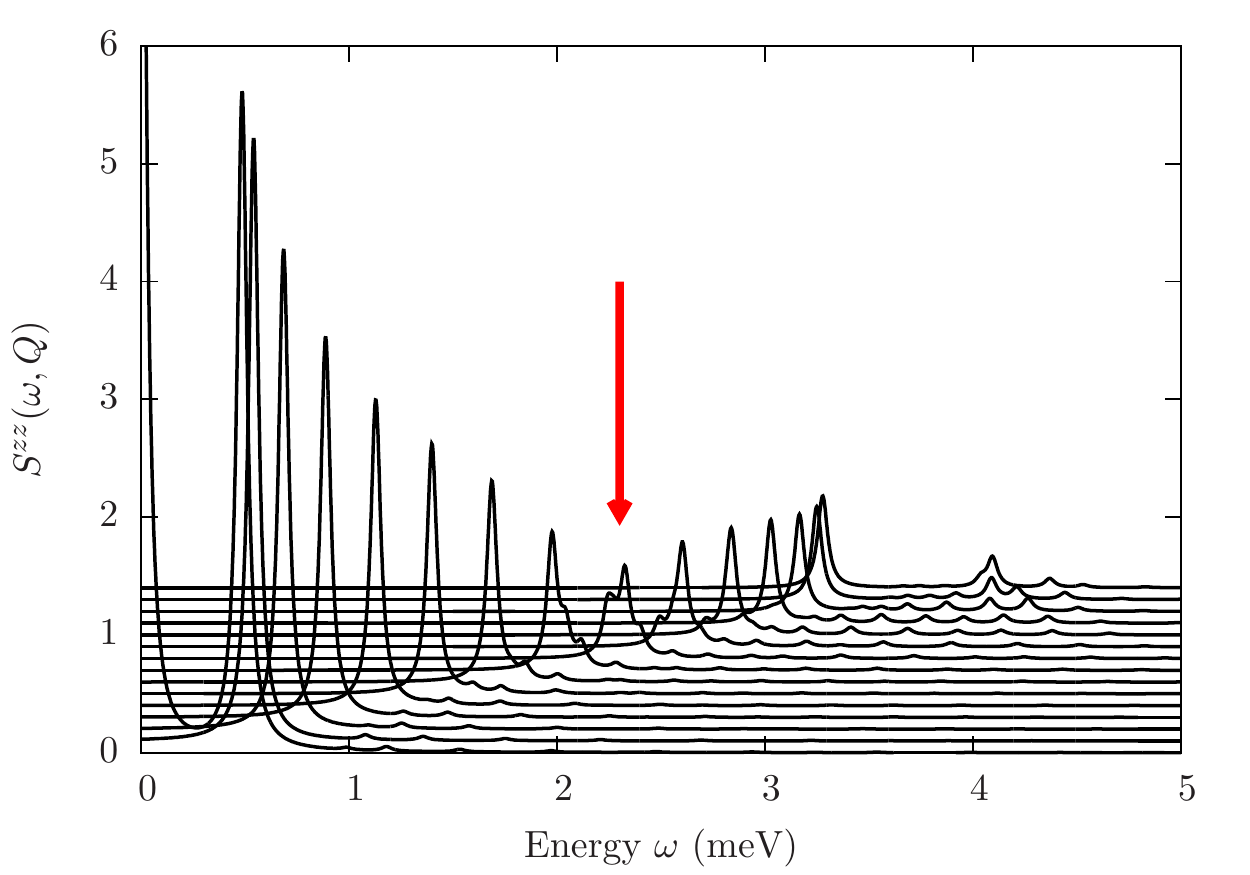} 
\end{tabular}
\caption{(Color online) Constant wave vector cuts ($l=0\to1$ in steps of $\delta l
= 1/14$, vertically displaced for clarity) of the dynamic structure factor
$S^{\alpha\alpha}(\omega,Q=l 2\pi/c)$ for (a) $\alpha = y$ and (b)
$\alpha=z$ for the $L=28$ site Hamiltonian~\fr{eq:pert} with
transverse field $B=7$~T and misalignment of $\theta\sim1.5^\circ$
($h_z = 0.062$~meV). We have used 120 Lanczos iterations in the
continued fraction and broadening parameter $\eta=0.01J$. The
arrow highlights the region of ``anomalous broadening'' of the
single particle mode at $\omega\approx 2-2.5$~meV. The corresponding
results for $h_z=0$ are shown in Fig.~\ref{fig:DSF}.}
\label{fig:L28_B7T}
\end{figure*}

Fig.~\ref{fig:L28_B7T} shows the Lanczos results for the
$\alpha=y,z$ components of the DSF in the $L=28$ chain at $B=7$~T
with a misalignment of $\theta\sim1.5^\circ$ ($h_z = 0.062$meV).
We see that when the single particle mode brushes the continuum
(at $\omega\approx 2-2.5$~meV, cf. Fig.~\ref{fig:Z2_Spectrum}) the
mode loses intensity and significantly broadens. This is consistent with the
range of momenta $l\approx0.5-0.7$ and frequency observed experimentally,
see Figs.~\ref{fig:Exp}(a),~\ref{fig:AnomBroad}(a) and~\ref{fig:cuts}(a).
We see that the multi-particle continuum feature at $E\approx 4$~meV, 
$l\approx 1$ persists, which is also consistent with experiment.

\begin{figure*}
\begin{tabular}{ll}
(a) & (b) \\ 
\includegraphics[width=0.48\textwidth]{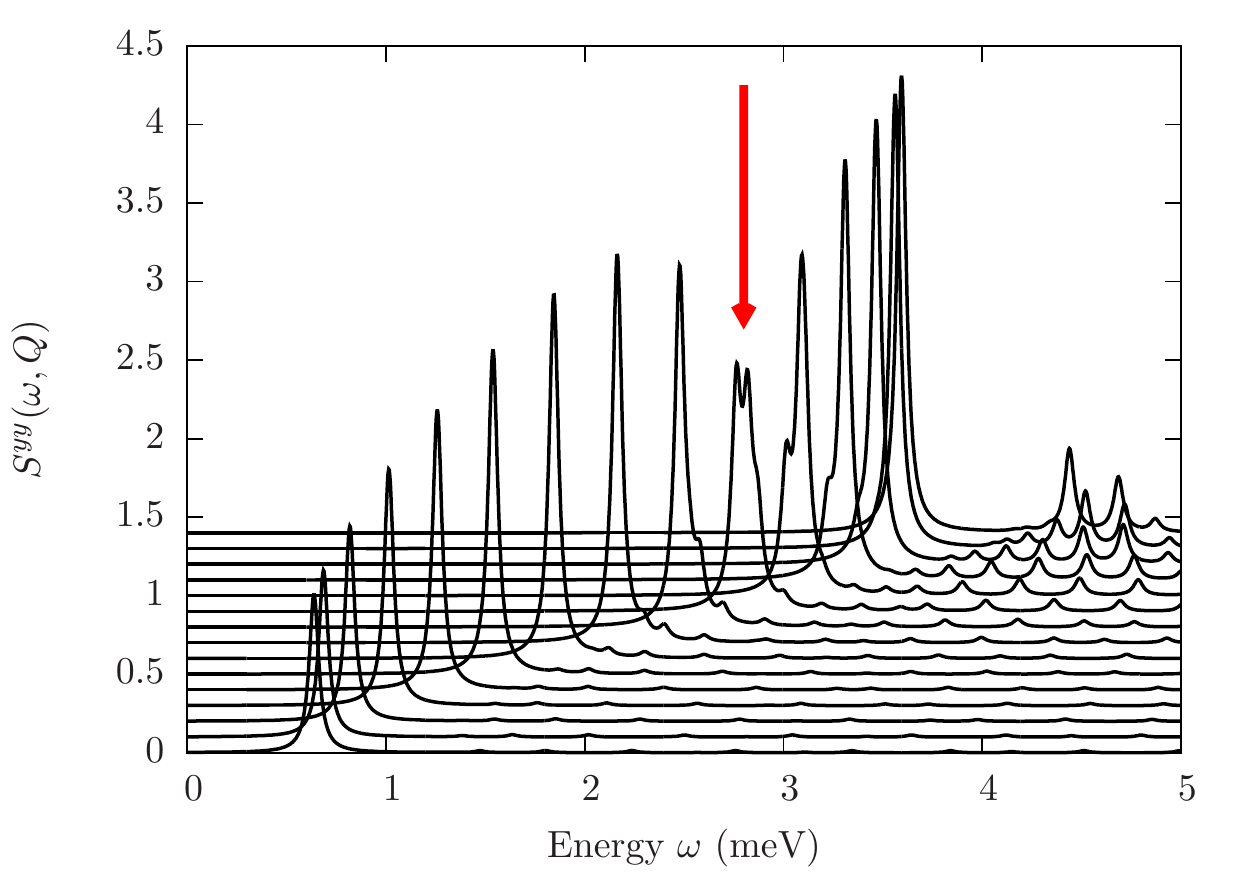} &
\includegraphics[width=0.48\textwidth]{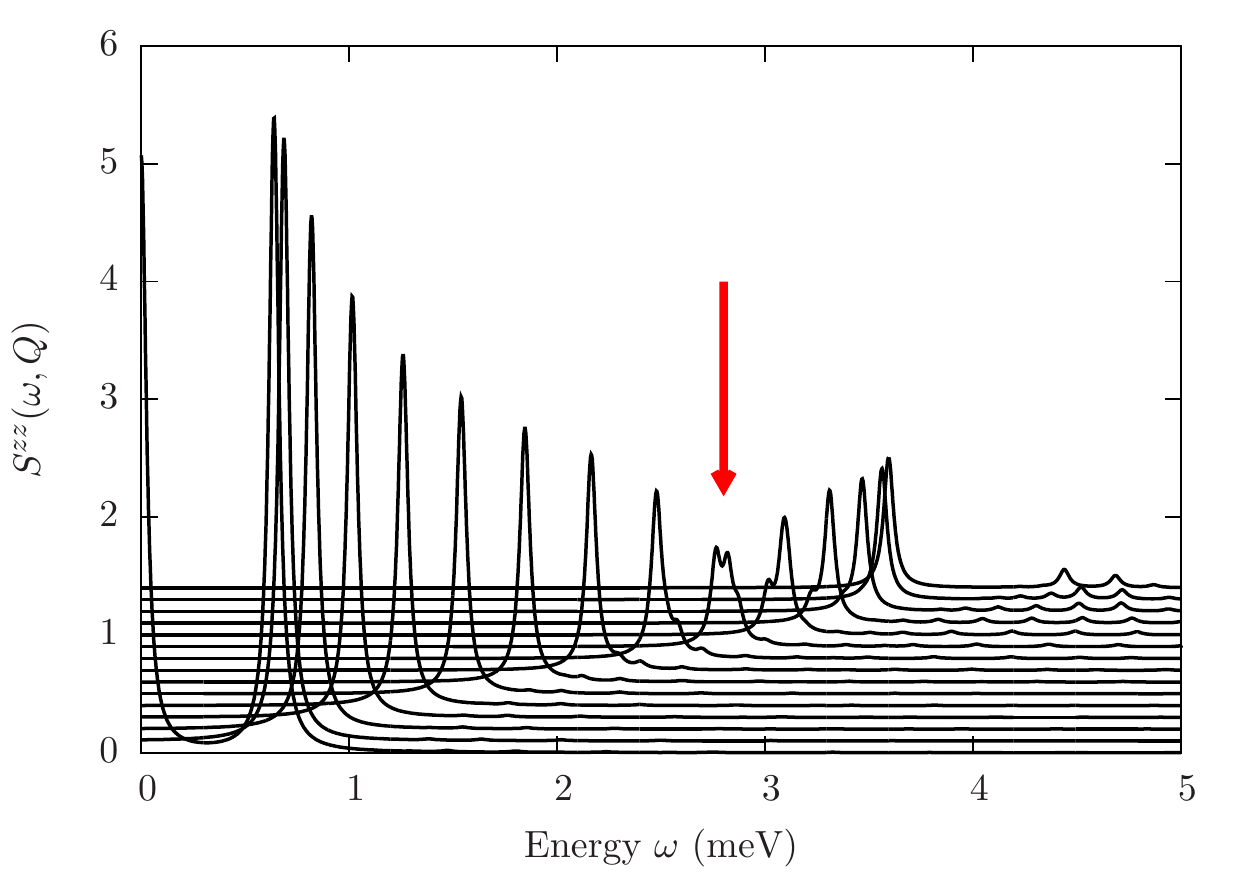} 
\end{tabular}
\caption{(Color online) Constant wave vector cuts ($l=0\to1$ in steps of $\delta l = 1/14$,
vertically displaced for clarity)
of the dynamic structure factor $S^{\alpha\alpha}(\omega,Q=l 2\pi/c)$ for
(a) $\alpha = y$ and (b) $\alpha=z$ for the $L=28$ site Hamiltonian~\fr{eq:pert} with
transverse field $B=8$~T and misalignment of $\theta\sim1.5^\circ$ ($h_z = 0.072$~meV).
We have used 120 Lanczos iterations in the continued fraction and
broadening parameter $\eta=0.01J$. The arrow highlights the region of
``anomalous broadening'' of the single particle mode at $\omega\approx 2.5-3.5$~meV.}
\label{fig:L28_B8T}
\end{figure*}

Analogous results for a field of $B=8$~T are shown
Fig.~\ref{fig:L28_B8T}. Compared to the $B=7$T data the region of
anomalous broadening has shifted slightly in energy and momentum
($l\approx0.55-0.75$) and the intensity loss is less pronounced,
reflecting the decreased overlap between the single particle mode
and the two-particle continuum, cf Fig.~\ref{fig:Z2_Spectrum}.
Note that the shift in energy and momentum and decreased loss of intensity 
is also observed in the data, see Figs.~\ref{fig:AnomBroad}(b) and~\ref{fig:cuts}(b).

\begin{figure*}
\begin{tabular}{ll}
(a) & (b) \\
\includegraphics[width=0.48\textwidth]{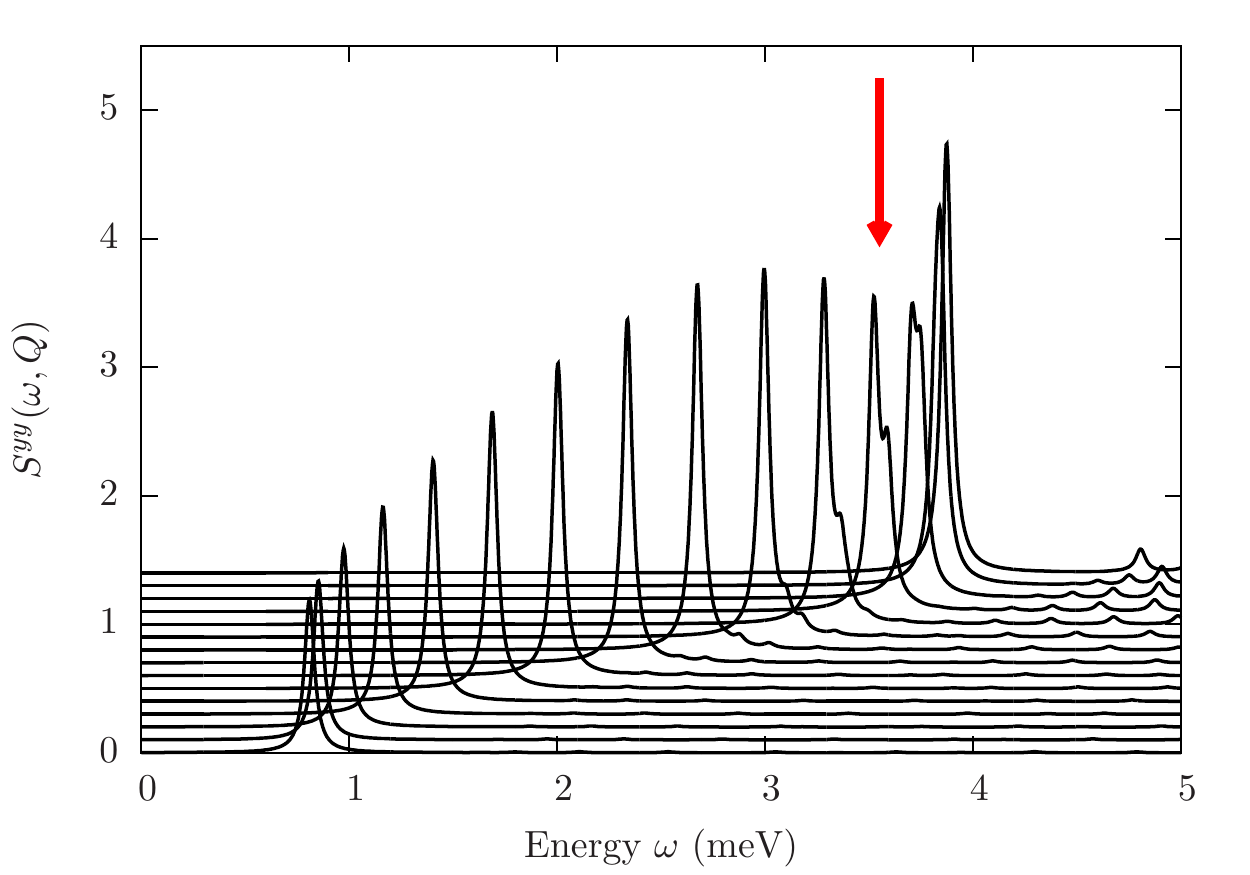}  &
\includegraphics[width=0.48\textwidth]{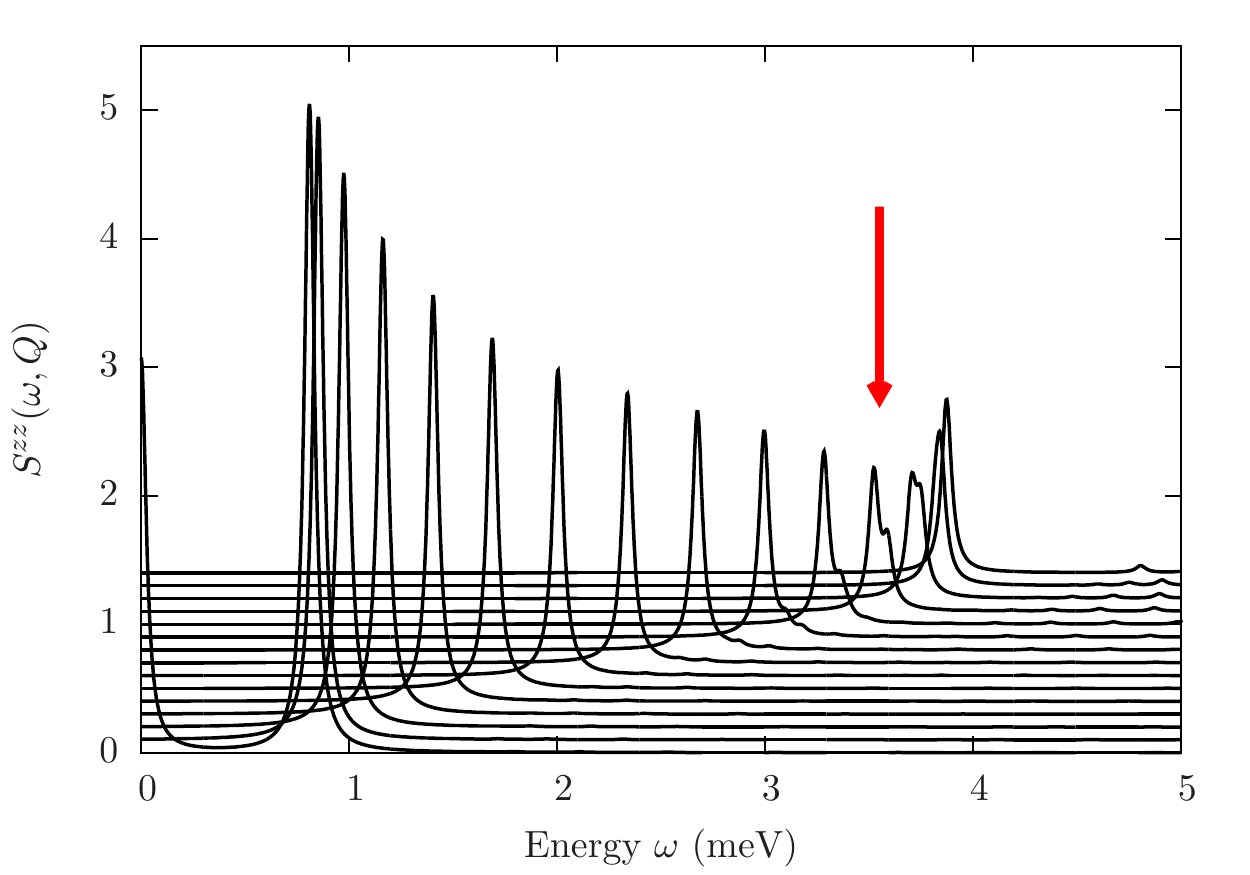} 
\end{tabular}
\caption{(Color online) Constant wave vector cuts ($l=0\to1$ in steps of $\delta l = 1/14$,
vertically displaced for clarity)
of the dynamic structure factor $S^{\alpha\alpha}(\omega,Q=l 2\pi/c)$ for
(a) $\alpha = y$ and (b) $\alpha=z$ for the $L=28$ site Hamiltonian~\fr{eq:pert} with
transverse field $B=9$~T and misalignment of $\theta\sim1.5^\circ$ ($h_z =
  0.081$~meV). We have used 120 Lanczos iterations in the continued fraction and
broadening parameter $\eta=0.01J$. The arrow highlights the broadening
region at $\omega \approx 3.5$~meV. }
\label{fig:L28_B9T}
\end{figure*}

The numerical calculations predict that upon increasing the field further to $B=9$T 
the anomalous broadening region shifts to wavevectors near $l\sim0.7$ and the 
broadening effect diminishes when compared to lower fields, compare Figs.~\ref{fig:L28_B8T} 
and Figs.~\ref{fig:L28_B9T}. The experimental data in Figs.~\ref{fig:AnomBroad}(a)--(c) indeed shows 
a shift with increasing field of the anomalous broadening region to higher energies along the dispersion 
bandwidth and to wavevectors further away from the $l=0.5$ zone boundary. However, the experimental
data also shows that the anomalous region at $B=9$T extends over a wider energy range and the 
broadening effect is more pronounced in the experimental data (Fig.~\ref{fig:AnomBroad}(c)) compared 
to the predictions of the theoretical model (Fig.~\ref{fig:L28_B9T}). There could be a number of possible 
reasons for these differences in detail.

Firstly, the misalignment angle could be dependent on the applied field.
This may be a result of the crystal not being completely rigid at high 
applied transverse fields. Whilst we have not extensively studied how 
the region of anomalous broadening moves with field-dependent misalignment, 
we have observed that increasing the longitudinal field at fixed transverse field 
results in the anomalous broadening becoming more severe and apparent over an
increased range of momenta. Secondly, there could be terms in the Hamiltonian 
beyond those taken into account in our minimal model~\fr{eq:model}. This can lead to 
the movement of the multi-particle continua in phase space, and as a result a change
in the region and severity of the anomalous broadening. Thirdly, the small system 
size $L=28$ in our exact diagonalization study may simply preclude an accurate 
description of the effect due to insufficient resolution in phase space or finite-size effects. 

\subsection{Quasi-particle breakdown}
Above we have shown that the addition of a small longitudinal
magnetic field component, consistent with small misalignment of
the crystal in experiment, leads to the broadening of the single
particle mode in the region $l\approx0.5-0.7$ and that this
broadening decreases with increased applied transverse field (for
fixed misalignment). High resolution inelastic neutron scattering
data in Figs.~\ref{fig:AnomBroad} and~\ref{fig:cuts} show that this 
indeed occurs in experiment, with the single particle mode becoming 
extremely broad and carrying little spectral weight around $l=0.5-0.65$. The level
of broadening observed in experiment is sufficient to say that the
quasi-particles are no longer \emph{well defined} over this region
of the Brillouin zone, a phenomena known as ``quasi-particle
breakdown''\cite{ZhitomirskyRMP13}.

A number of mechanisms for quasi-particle breakdown (and
specifically ``spontaneous magnon decay'' in quantum magnets) are
discussed in
Refs.~\onlinecite{ZhitomirskyRMP13,FischerDiss,ZhitomirskyPRB06,KolezhukPRL06,
BibikovPRB07,FischerNewJPhys10,FischerEPL11}, including the case
of field-induced decay. Most experimental observations of
quasi-particle breakdown have so far been limited to the case
where the single particle mode enters the two-particle continuum
and terminates, such as in quasi-2D quantum
magnets\cite{StoneNature06} and quasi-1D spin-1
chains\cite{MasudaPRL06}.

In this case we observe something more unusual: two region of the
Brillouin zone (0$\le | l | \lesssim 0.5$ and $0.7\lesssim| l| \le
1$) have coherent well-defined single particle excitations, whilst
in the intermediate region $0.5\lesssim|l|\lesssim0.7$
quasi-particle breakdown occurs. For the smallest fields that we
examine ($B=7$~T) this effect is particularly severe in experiments
(see Figs.~\ref{fig:AnomBroad}(a) and~\ref{fig:cuts}(a)), where one could easily believe
that a gap has opened in the single particle dispersion. Compare
this to a similar field-tuned effect seen in the quasi-2D quantum
magnet Ba$_2$MnGe$_2$O$_7$, where the excitation is broadened, but
without the severe loss of intensity~\cite{MasudaPRB10}.

The quasi-particle breakdown in \CoNbO\ is a direct result of
\emph{explicit symmetry breaking within the experimental setting}, and
highlights the crucial role that symmetry breaking perturbations can
play.

\section{Conclusions}
\label{sec:Conclusion}

Motivated by recent inelastic neutron scattering
experiments\cite{CabreraArxiv14}, we have investigated the origin
of the anomalous broadening of the single particle dispersion in
the quasi-one-dimensional ferromagnet \CoNbO. We have presented
high-resolution inelastic neutron scattering data (see
Fig.~\ref{fig:AnomBroad}) showing that the observed anomalous
broadening has a non-trivial field dependence and is particularly
severe at the small transverse field strengths (7~T), where the
broadening may easily be mistaken for a gap in the single particle
dispersion. To understand this behaviour, we have proposed a
one-dimensional spin Hamiltonian whose parameters we fix by
fitting the single particle dispersion to inelastic neutron scattering data
presented in Fig.~\ref{fig:Exp}.

Having fixed the exchange parameters of our effective model, we add a
\emph{single} free parameter to our model -- a longitudinal magnetic
field. Such an addition is entirely reasonable, as we expect a small
longitudinal field to arise from slight misalignment of the crystal in
experiment. Crucially, this longitudinal field breaks spin inversion
symmetry ($S^z \to -S^z$) which forbids transitions between the
one-particle mode and the two-particle continuum. The breaking of this
symmetry has a profound effect on the dynamical structure factor of
the quantum spin model -- in regions of the Brillouin zone where the
two-particle continuum overlaps with the single particle mode (see
Fig.~\ref{fig:Z2_Spectrum}) we see that the single particle mode loses
weight and broadens (see Figs.~\ref{fig:L28_B7T}~and~\ref{fig:L28_B8T}
for exact diagonalization data). This broadening occurs due to the
longitudinal field inducing the spontaneous decay of the single
particle excitation into multi-particle  excitations, an example of
``quasi-particle breakdown''\cite{ZhitomirskyRMP13}. \CoNbO\ is
particularly unusual in this regard as the region of quasi-particle
breakdown separates two regions of coherent quasi-particles in the
Brillouin zone.

\textit{Acknowledgements}.
This work was supported by the EPSRC under Grants No. EP/I032487/1
(FHLE and NJR) and EP/H014934/1 (RC and IC).

\onecolumngrid
\appendix
\section{Transforming the spin Hamiltonian~\fr{eq:model} into the fermion Hamiltonian~\fr{eq:Bog_Ham}}
\label{App:SelfConBog}

Starting from the Hamiltonian~\fr{eq:model}, we start by rotating
the spin quantization axes by $\pi/2$ about $S^y$ to be in keeping
with standard conventions. We then perform a Jordan-Wigner
transformation and subsequently Fourier transform the resulting
fermionic theory to obtain the momentum space Hamiltonian $H= H_0 + H_{\rI} + E_0$ where $E_0$ is an
additive constant that rescales the absolute energy and is
neglected herein, $H_0$ contains only fermion bilinears and
$H_{\rm int}$ is quartic in the fermion operators \bea H_0 &=&
\frac{1}{2} \sum_k \Big(c^\dagger_k~~c_{-k}\Big) \left(
\begin{array}{cc}
A_k & iB_k \\ -iB_k & -A_k \end{array}\right) \left(\begin{array}{c} c_k \\ c_{-k}^\dagger \end{array} \right),\nn
H_{\rI} &=& \frac{J}{2L}(\lm_1-\lm_3) \sum_{k_i} \Big[ f_{(k_1,k_2,k_3)(k_4)}c_{k_1}^\dagger c_{k_2}^\dagger
c_{k_3}^\dagger c_{-k_4} +{\rm H.c.} \Big]\nn
&-&  \frac{J}{2L} \sum_{k_i} \Big[2\lm_2 h_{(k_1,k_2)(k_3,k_4)} +  2\lm_3 h_{(2k_1,2k_2)(2k_3,2k_4)}
+ (\lm_1+\lm_3)g_{(k_1,k_2)(k_3,k_4)} \Big] c^\dagger_{k_1}c^\dagger_{k_2} c_{-k_3} c_{-k_4},\nonumber
\eea
The matrix elements of $H_0$ are given by
\bea
A_k &=& \frac{J}{2}(1+\lm_2)\cos(k) + \frac{J}{2}(\lm_1+\lm_3)\cos(2k) + h_x - J(\lm_2 +\lm_3),\nn
B_k &=& -\frac{J}{2}(1-\lm_2)\sin(k) - \frac{J}{2}(\lm_1-\lm_3)\sin(2k),\nonumber
\eea
whilst the vertex factors appearing in $H_\rI$ take the form
\bea
f_{(k_1,k_2,k_3)(k_4)} &=& \frac{i}{3}\left[ \sin(k_3-k_1) + \sin(k_1-k_2) + \sin(k_2-k_3)\right]\delta_{\sum_j k_j,0},\nn
g_{(k_1,k_2)(k_3,k_4)} &=& \frac{1}{2}\left[\cos(k_4-k_1)-\cos(k_4-k_2)+\cos(k_3-k_2)-\cos(k_3-k_1)\right]\delta_{\sum_j k_j,0},\nn
h_{(k_1,k_2)(k_3,k_4)} &=& \frac{1}{4}\left[\cos(k_1+k_3)-\cos(k_2+k_3)+\cos(k_2+k_4)-\cos(k_1+k_4)\right]\delta_{\sum_j k_j,0}, \nonumber
\eea
which are antisymmetric under pair-wise exchange of indices appearing within the same brackets $(\ldots)$
and impose momentum conservation.

We now diagonalize the quadratic part of the Hamiltonian by performing a self-consistent
Bogoliubov transformation. We define the Bogoliubov fermions $a_k$ by
\bea
c^\dagger_{k} = -i\cos\theta_k a^\dagger_k - \sin\theta_k a_{-k},\qquad
c_{k} = i\cos\theta_k a_k - \sin\theta_k a^\dagger_{-k},
\label{eq:bogtrans}
\eea
where the Bogoliubov parameter $\theta_{k}=-\theta_{-k}$ satisfies the
self-consistency condition $A_k \sin(2\theta_k) - B_k\cos(2\theta_k) = 0$. The
quadratic part of the Hamiltonian then becomes diagonal
\be
H_0= \frac{1}{2} \sum_k \Big(a^\dagger_k~~a_{-k}\Big) \left( \begin{array}{cc}
\sqrt{A_k^2+B_k^2}  & 0 \\ 0 & -\sqrt{A_k^2+B_k^2}\end{array}\right) \left(\begin{array}{c} a_k \\ a_{-k}^\dagger \end{array} \right).
\label{eq:bogquad}
\ee

Let us now consider the action of the Bogoliubov
transformation~\fr{eq:bogtrans} on the interaction term of the
Hamiltonian $H_\rI$. It is clear that many of the transformed
terms in $H_\rI$ will not be normal ordered. The normal ordering
of these terms will generate fermion bilinear terms that
contribute to both the diagonal and off-diagonal elements of $H_0$
in Eq.~\fr{eq:bogquad}. In order that the quadratic part of the
Hamiltonian is diagonal, we impose a self-consistency condition on
the Bogoliubov parameter: it must be chosen such that the
off-diagonal terms that result from normal-ordering interaction
terms vanish. The resulting self-consistency condition for the
Bogoliubov parameter is \bea &&\bigg[A_k+\sum_q\Theta_1(k,q)
\bigg]\sin2\theta_k -\bigg[B_k +\sum_q
\Theta_2(k,q)\bigg]\cos2\theta_k  = 0,
 \label{eq:selfcon}
\eea
where we have defined the functions
\bea
\Theta_1(k,q) &=& - \frac{4J}{L} \bigg[\frac{1}{2} (\lambda_1+\lambda_3)g_{(k,q)(-q,-k)}+
\lambda_2 h_{(k,q)(-q,-k)} + \lambda_3 h_{(2k,2q)(-2q,-2k)} \bigg] \sin^2\theta_q\nn
&& +  \frac{3J}{2L} (\lambda_1-\lambda_3) if_{(k,q,-q)(-k)}\sin2\theta_q,\nn
\Theta_{2}(k,q)&=& \frac{J}{L} \bigg[\frac{1}{2}(\lambda_1+\lambda_3)g_{(k,-k)(q,-q)}
+\lambda_2 h_{(k,-k)(q,-q)} + \lambda_3 h_{(2k,-2k)(2q,-2q)} \bigg] \sin2\theta_q\nn
&& + \frac{3J}{L} (\lambda_1-\lambda_3) if_{(k,q,-k)(-q)}\sin^2\theta_q, \nonumber
\eea
which also depend upon the Bogoliubov parameter.

The self-consistency condition~\fr{eq:selfcon} perturbatively
modifies the Bogoliubov parameter. Due to the complicated
structure Eq.~\fr{eq:selfcon}, we solve the set of non-linear
simultaneous equations numerically using standard techniques.
Following the imposition of the self-consistency condition, we
obtain the Hamiltonian \fr{eq:Bog_Ham} with dispersion
relation~\fr{SelfConE}.

\section{Vertex functions}
\label{app:Vertex}

The vertex functions $V_0,~V_1,~V_2$ in Eq.~\fr{eq:Bog_Ham} are obtained by normal-ordering of the
four-fermion terms after Bogoliubov transformation. By symmetry, they can be expressed in terms of
summations over permutations of indices. For example
\bea
V_0(k_1,k_2,k_3,k_4) &=& \delta_{\sum_j k_j,0} \Bigg\{\frac{1}{96}\left( \lm_3-\lm_1\right) \sum_{P\in S_4} {\rm sgn}(P)
\cos\bigg[ k_{P_1} - k_{P_2} + \theta_{k_{P_1}} + \theta_{k_{P_2}} + \theta_{k_{P_3}} - \theta_{k_{P_4}}\bigg] \nn
&&+ \frac{1}{96}\sum_{j=2}^3 \lm_j \sum_{P\in S_4} {\rm sgn}(P)
\cos\bigg[(j-1)\Big(k_{P_1} + k_{P_2}\Big) + \theta_{k_{P_1}} - \theta_{k_{P_2}} + \theta_{k_{P_3}} - \theta_{k_{P_4}}\bigg]\Bigg\}\ ,
\nonumber
\eea
where the permutation $P$ acts on the set $P:\{1,2,3,4\} \to \{P_1,P_2,P_3,P_4\}$.

The vertex which changes quasi-particle number by two is given by
\bea
V_1(k_1,k_2,k_3,k_4) = \delta_{\sum_j k_j,0} \bigg[ V_1^{(12)}(k_1,k_2,k_3,k_4) + V_1^{(23)}(k_1,k_2,k_3,k_4) + V_1^{(13)}(k_1,k_2,k_3,k_4) \bigg],
\nonumber
\eea
where
\bea
V_1^{(12)}(k_1,k_2,k_3,k_4) &=& \frac{i}{24}\left(\lm_1-\lm_2 \right) \sum_{Q\in S_3} {\rm sgn}(Q)
\Bigg\{ \sin\bigg[k_{Q_1} - k_4 + \theta_{k_{Q_1}}-\theta_{k_{Q_2}} + \theta_{k_{Q_3}} - \theta_{k_4} \bigg]\nn
&&\hspace{3.8cm}
+\sin\bigg[k_{Q_1} - k_4 - \theta_{k_{Q_1}}+\theta_{k_{Q_2}} - \theta_{k_{Q_3}} + \theta_{k_4} \bigg]\nn
&&\hspace{3.8cm}
-\sin\bigg[k_{Q_1} - k_{Q_2} + \theta_{k_{Q_1}}-\theta_{k_{Q_2}} - \theta_{k_{Q_3}} + \theta_{k_4} \bigg]\nn
&&\hspace{3.8cm}
-\sin\bigg[k_{Q_1} - k_{Q_2} - \theta_{k_{Q_1}}+\theta_{k_{Q_2}} - \theta_{k_{Q_3}} + \theta_{k_4} \bigg]
\Bigg\},\nonumber
\eea
\bea
V_1^{(23)}(k_1,k_2,k_3,k_4)&=&-\frac{i}{24} \sum_{j=2}^3 \lm_j \sum_{Q\in S_3} {\rm sgn}(Q) \Bigg\{
\sin\bigg[(j-1)\Big(k_{Q_1} + k_{Q_2}\Big) - \theta_{k_{Q_1}}+\theta_{k_{Q_2}} - \theta_{k_{Q_3}} + \theta_{k_4} \bigg]\nn
&&\hspace{3.6cm}+
\sin\bigg[(j-1)\Big(k_{Q_1} + k_{Q_2} \Big) - \theta_{k_{Q_1}}+\theta_{k_{Q_2}} + \theta_{k_{Q_3}} - \theta_{k_4} \bigg]\nn
&&\hspace{3.6cm}+
\sin\bigg[(j-1)\Big(k_{Q_1} + k_{4}\Big) + \theta_{k_{Q_1}}-\theta_{k_{Q_2}} + \theta_{k_{Q_3}} - \theta_{k_4} \bigg]\nn
&&\hspace{3.6cm}+
\sin\bigg[(j-1)\Big(k_{Q_1} + k_{4}\Big) - \theta_{k_{Q_1}}-\theta_{k_{Q_2}} + \theta_{k_{Q_3}} + \theta_{k_4} \bigg]\Bigg\},\nonumber
\eea
\bea
V_1^{(13)}(k_1,k_2,k_3,k_4)&=& \frac{i}{24} (\lm_1+\lm_3) \sum_{Q\in S_3}{\rm sgn}(Q) \Bigg\{
 \sin\bigg[k_{Q_1} - k_{Q_2} - \theta_{k_{Q_1}}+\theta_{k_{Q_2}} + \theta_{k_{Q_3}} - \theta_{k_4} \bigg]\nn
&&\hspace{3.8cm}
-\sin\bigg[k_{Q_1} - k_{Q_2} + \theta_{k_{Q_1}}-\theta_{k_{Q_2}} - \theta_{k_{Q_3}} + \theta_{k_4} \bigg]\nn
&&\hspace{3.8cm}
+ \sin\bigg[k_{Q_1} - k_4 + \theta_{k_{Q_1}}-\theta_{k_{Q_2}} + \theta_{k_{Q_3}} - \theta_{k_4} \bigg]\nn
&&\hspace{3.8cm}
-\sin\bigg[k_{Q_1} - k_4 - \theta_{k_{Q_1}}+\theta_{k_{Q_2}} - \theta_{k_{Q_3}} + \theta_{k_4} \bigg]\Bigg\}.\nonumber
\eea
Here in $V_1^{(12)}$, $V_1^{(23)}$ and $V_1^{(13)}$ the permutation $Q$ acts on the set $Q:\{ 1, 2, 3 \}\to\{Q_1,Q_2,Q_3\}$.

The remaining vertex function that preserves quasiparticle number is given by
\bea
V_2(k_1,k_2,k_3,k_4) = \delta_{\sum_j k_j , 0}\bigg[V_2^{(1)}(k_1,k_2,k_3,k_4) + V_2^{(23)}(k_1,k_2,k_3,k_4) + V_2^{(3)}(k_1,k_2,k_3,k_4)\bigg],
\nonumber
\eea
with
\bea
V^{(1)}_2(k_1,k_2,k_3,k_4) &=& \frac{\lm_1}{4} \sum_{P,Q \in S_2}
{\rm sgn}(P){\rm sgn}(Q) \Bigg\{
\cos\bigg[k_{P_1}- k_{P_2} + \theta_{k_1}-\theta_{k_2} + \theta_{k_{Q_3}} - \theta_{k_{Q_4}}\bigg]\nn
&&\hspace{3.7cm}+
\cos\bigg[k_{Q_3} - k_{Q_4} + \theta_{k_3}-\theta_{k_4} + \theta_{k_{P_1}} - \theta_{k_{P_2}}\bigg]\nn
&&\hspace{3.7cm}+
\cos\bigg[k_{P_1} - k_{Q_3} + \theta_{k_{P_1}}+\theta_{k_{P_2}} - \theta_{k_{Q_3}} - \theta_{k_{Q_4}}\bigg]\nn
&&\hspace{3.7cm}+
\cos\bigg[k_{P_1} - k_{Q_3} + \theta_{k_{P_1}}-\theta_{k_{P_2}} - \theta_{k_{Q_3} }+ \theta_{k_{Q_4}}\bigg]
\Bigg\},\nonumber
\eea
\bea
V^{(23)}(k_1,k_2,k_3,k_4) &=& \frac{1}{8}\sum_{j=2}^3 \lm_j \sum_{P,Q \in S_2} {\rm sgn}(P){\rm sgn}(Q) \Bigg\{
\cos\bigg[(j-1)\Big(k_{P_1}+ k_{P_2}\Big) + \theta_{k_1}-\theta_{k_2} - \theta_{k_{Q_3}} + \theta_{k_{Q_4}}\bigg]\nn
&&\hspace{4.5cm}+
\cos\bigg[(j-1)\Big(k_{Q_3}+ k_{Q_4}\Big) + \theta_{k_3}-\theta_{k_4} - \theta_{k_{P_1}} + \theta_{k_{P_2}}\bigg]\nn
&&\hspace{4.5cm}-
\cos\bigg[(j-1)\Big(k_{P_1}+ k_{Q_3}\Big) + \theta_{k_{P_1}}+\theta_{k_{P_2}} - \theta_{k_{Q_3}} - \theta_{k_{Q_4}}\bigg]\nn
&&\hspace{4.5cm}-
\cos\bigg[(j-1)\Big(k_{P_1} + k_{Q_3}\Big) + \theta_{k_{P_1}}-\theta_{k_{P_2}} - \theta_{k_{Q_3}} + \theta_{k_{Q_4}}\bigg]\nn
&&\hspace{4.5cm}-
\cos\bigg[(j-1)\Big(k_{P_1} + k_{Q_3}\Big) - \theta_{k_{P_1}}-\theta_{k_{P_2}} + \theta_{k_{Q_3}} + \theta_{k_{Q_4}}\bigg]\nn
&&\hspace{4.5cm}-
\cos\bigg[(j-1)\Big(k_{P_1} + k_{Q_3}\Big) - \theta_{k_{P_1}}+\theta_{k_{P_2}} + \theta_{k_{Q_3}} - \theta_{k_{Q_4}}\bigg]\Bigg\},\nonumber
\eea
\bea
V_2^{(3)}(k_1,k_2,k_3,k_4) &=& \frac{\lm_3}{4} \sum_{P,Q \in S_2} {\rm sgn}(P){\rm sgn}(Q) \Bigg\{
\cos\bigg[k_{P_1} - k_{Q_3} - \theta_{k_{P_1}}+\theta_{k_{P_2}} + \theta_{k_{Q_3}} - \theta_{k_{Q_4}}\bigg]\nn
&&\hspace{3.7cm}+
\cos\bigg[k_{P_1} - k_{Q_3} - \theta_{k_{P_1}}-\theta_{k_{P_2}} + \theta_{k_{Q_3}} + \theta_{k_{Q_4}}\bigg]\Bigg\},\nonumber
\eea
where $P$ is the permutation acting on the set $P:\{ 1, 2 \}\to\{P_1,P_2\}$ and the permutation $Q$ acts on the set $Q:\{3, 4\}\to\{Q_3,Q_4\}$.

\twocolumngrid
\bibliography{bib}

\begin{thebibliography}{52}%
\makeatletter
\providecommand \@ifxundefined [1]{%
 \@ifx{#1\undefined}
}%
\providecommand \@ifnum [1]{%
 \ifnum #1\expandafter \@firstoftwo
 \else \expandafter \@secondoftwo
 \fi
}%
\providecommand \@ifx [1]{%
 \ifx #1\expandafter \@firstoftwo
 \else \expandafter \@secondoftwo
 \fi
}%
\providecommand \natexlab [1]{#1}%
\providecommand \enquote  [1]{``#1''}%
\providecommand \bibnamefont  [1]{#1}%
\providecommand \bibfnamefont [1]{#1}%
\providecommand \citenamefont [1]{#1}%
\providecommand \href@noop [0]{\@secondoftwo}%
\providecommand \href [0]{\begingroup \@sanitize@url \@href}%
\providecommand \@href[1]{\@@startlink{#1}\@@href}%
\providecommand \@@href[1]{\endgroup#1\@@endlink}%
\providecommand \@sanitize@url [0]{\catcode `\\12\catcode `\$12\catcode
  `\&12\catcode `\#12\catcode `\^12\catcode `\_12\catcode `\%12\relax}%
\providecommand \@@startlink[1]{}%
\providecommand \@@endlink[0]{}%
\providecommand \url  [0]{\begingroup\@sanitize@url \@url }%
\providecommand \@url [1]{\endgroup\@href {#1}{\urlprefix }}%
\providecommand \urlprefix  [0]{URL }%
\providecommand \Eprint [0]{\href }%
\providecommand \doibase [0]{http://dx.doi.org/}%
\providecommand \selectlanguage [0]{\@gobble}%
\providecommand \bibinfo  [0]{\@secondoftwo}%
\providecommand \bibfield  [0]{\@secondoftwo}%
\providecommand \translation [1]{[#1]}%
\providecommand \BibitemOpen [0]{}%
\providecommand \bibitemStop [0]{}%
\providecommand \bibitemNoStop [0]{.\EOS\space}%
\providecommand \EOS [0]{\spacefactor3000\relax}%
\providecommand \BibitemShut  [1]{\csname bibitem#1\endcsname}%
\let\auto@bib@innerbib\@empty
\bibitem [{\citenamefont {Cabrera}\ \emph {et~al.}(2014)\citenamefont
  {Cabrera}, \citenamefont {Thompson}, \citenamefont {Coldea}, \citenamefont
  {Prabhakaran}, \citenamefont {Bewley}, \citenamefont {Guidi}, \citenamefont
  {Rodriguez-Rivera},\ and\ \citenamefont {Stock}}]{CabreraArxiv14}%
  \BibitemOpen
  \bibfield  {author} {\bibinfo {author} {\bibfnamefont {I.}~\bibnamefont
  {Cabrera}}, \bibinfo {author} {\bibfnamefont {J.~D.}\ \bibnamefont
  {Thompson}}, \bibinfo {author} {\bibfnamefont {R.}~\bibnamefont {Coldea}},
  \bibinfo {author} {\bibfnamefont {D.}~\bibnamefont {Prabhakaran}}, \bibinfo
  {author} {\bibfnamefont {R.~I.}\ \bibnamefont {Bewley}}, \bibinfo {author}
  {\bibfnamefont {T.}~\bibnamefont {Guidi}}, \bibinfo {author} {\bibfnamefont
  {J.~A.}\ \bibnamefont {Rodriguez-Rivera}}, \ and\ \bibinfo {author}
  {\bibfnamefont {C.}~\bibnamefont {Stock}},\ }\href {\doibase
  10.1103/PhysRevB.90.014418} {\bibfield  {journal} {\bibinfo  {journal} {Phys.
  Rev. B}\ }\textbf {\bibinfo {volume} {90}},\ \bibinfo {pages} {014418}
  (\bibinfo {year} {2014})}\BibitemShut {NoStop}%
\bibitem [{\citenamefont {Anderson}(1952)}]{AndersonPR52}%
  \BibitemOpen
  \bibfield  {author} {\bibinfo {author} {\bibfnamefont {P.~W.}\ \bibnamefont
  {Anderson}},\ }\href {\doibase 10.1103/PhysRev.86.694} {\bibfield  {journal}
  {\bibinfo  {journal} {Phys. Rev.}\ }\textbf {\bibinfo {volume} {86}},\
  \bibinfo {pages} {694} (\bibinfo {year} {1952})}\BibitemShut {NoStop}%
\bibitem [{\citenamefont {Kubo}(1952)}]{KuboPR52}%
  \BibitemOpen
  \bibfield  {author} {\bibinfo {author} {\bibfnamefont {R.}~\bibnamefont
  {Kubo}},\ }\href {\doibase 10.1103/PhysRev.87.568} {\bibfield  {journal}
  {\bibinfo  {journal} {Phys. Rev.}\ }\textbf {\bibinfo {volume} {87}},\
  \bibinfo {pages} {568} (\bibinfo {year} {1952})}\BibitemShut {NoStop}%
\bibitem [{\citenamefont {Kubo}(1953)}]{KuboRMP53}%
  \BibitemOpen
  \bibfield  {author} {\bibinfo {author} {\bibfnamefont {R.}~\bibnamefont
  {Kubo}},\ }\href {\doibase 10.1103/RevModPhys.25.344} {\bibfield  {journal}
  {\bibinfo  {journal} {Rev. Mod. Phys.}\ }\textbf {\bibinfo {volume} {25}},\
  \bibinfo {pages} {344} (\bibinfo {year} {1953})}\BibitemShut {NoStop}%
\bibitem [{\citenamefont {Dyson}(1956{\natexlab{a}})}]{DysonPR56a}%
  \BibitemOpen
  \bibfield  {author} {\bibinfo {author} {\bibfnamefont {F.~J.}\ \bibnamefont
  {Dyson}},\ }\href {\doibase 10.1103/PhysRev.102.1217} {\bibfield  {journal}
  {\bibinfo  {journal} {Phys. Rev.}\ }\textbf {\bibinfo {volume} {102}},\
  \bibinfo {pages} {1217} (\bibinfo {year} {1956}{\natexlab{a}})}\BibitemShut
  {NoStop}%
\bibitem [{\citenamefont {Dyson}(1956{\natexlab{b}})}]{DysonPR56b}%
  \BibitemOpen
  \bibfield  {author} {\bibinfo {author} {\bibfnamefont {F.~J.}\ \bibnamefont
  {Dyson}},\ }\href {\doibase 10.1103/PhysRev.102.1230} {\bibfield  {journal}
  {\bibinfo  {journal} {Phys. Rev.}\ }\textbf {\bibinfo {volume} {102}},\
  \bibinfo {pages} {1230} (\bibinfo {year} {1956}{\natexlab{b}})}\BibitemShut
  {NoStop}%
\bibitem [{\citenamefont {Van~Kranendonk}\ and\ \citenamefont
  {Van~Vleck}(1958)}]{VanKranendonkRMP58}%
  \BibitemOpen
  \bibfield  {author} {\bibinfo {author} {\bibfnamefont {J.}~\bibnamefont
  {Van~Kranendonk}}\ and\ \bibinfo {author} {\bibfnamefont {J.~H.}\
  \bibnamefont {Van~Vleck}},\ }\href {\doibase 10.1103/RevModPhys.30.1}
  {\bibfield  {journal} {\bibinfo  {journal} {Rev. Mod. Phys.}\ }\textbf
  {\bibinfo {volume} {30}},\ \bibinfo {pages} {1} (\bibinfo {year}
  {1958})}\BibitemShut {NoStop}%
\bibitem [{\citenamefont {Akhiezer}\ \emph {et~al.}(1958)\citenamefont
  {Akhiezer}, \citenamefont {Bar’yakhtar},\ and\ \citenamefont
  {Peletminskii}}]{AkhiezerBook}%
  \BibitemOpen
  \bibfield  {author} {\bibinfo {author} {\bibfnamefont {A.~I.}\ \bibnamefont
  {Akhiezer}}, \bibinfo {author} {\bibfnamefont {V.~G.}\ \bibnamefont
  {Bar’yakhtar}}, \ and\ \bibinfo {author} {\bibfnamefont {S.~V.}\
  \bibnamefont {Peletminskii}},\ }\href@noop {} {\emph {\bibinfo {title} {Spin
  Waves}}}\ (\bibinfo  {publisher} {North-Holland},\ \bibinfo {year}
  {1958})\BibitemShut {NoStop}%
\bibitem [{\citenamefont {Mattis}(1981)}]{MattisBook}%
  \BibitemOpen
  \bibfield  {author} {\bibinfo {author} {\bibfnamefont {D.~C.}\ \bibnamefont
  {Mattis}},\ }\href@noop {} {\emph {\bibinfo {title} {The Theory of
  Magnetism}}}\ (\bibinfo  {publisher} {Springer},\ \bibinfo {year}
  {1981})\BibitemShut {NoStop}%
\bibitem [{\citenamefont {Pauthenet}(1982)}]{PauthenetJAP82}%
  \BibitemOpen
  \bibfield  {author} {\bibinfo {author} {\bibfnamefont {R.}~\bibnamefont
  {Pauthenet}},\ }\href {\doibase 10.1063/1.330287} {\bibfield  {journal}
  {\bibinfo  {journal} {Journal of Applied Physics}\ }\textbf {\bibinfo
  {volume} {53}},\ \bibinfo {pages} {8187} (\bibinfo {year}
  {1982})}\BibitemShut {NoStop}%
\bibitem [{\citenamefont {Manousakis}(1991)}]{ManousakisRMP91}%
  \BibitemOpen
  \bibfield  {author} {\bibinfo {author} {\bibfnamefont {E.}~\bibnamefont
  {Manousakis}},\ }\href {\doibase 10.1103/RevModPhys.63.1} {\bibfield
  {journal} {\bibinfo  {journal} {Rev. Mod. Phys.}\ }\textbf {\bibinfo {volume}
  {63}},\ \bibinfo {pages} {1} (\bibinfo {year} {1991})}\BibitemShut {NoStop}%
\bibitem [{\citenamefont {Zhitomirsky}\ and\ \citenamefont
  {Chernyshev}(1999)}]{ZhitomirskyPRL99}%
  \BibitemOpen
  \bibfield  {author} {\bibinfo {author} {\bibfnamefont {M.~E.}\ \bibnamefont
  {Zhitomirsky}}\ and\ \bibinfo {author} {\bibfnamefont {A.~L.}\ \bibnamefont
  {Chernyshev}},\ }\href {\doibase 10.1103/PhysRevLett.82.4536} {\bibfield
  {journal} {\bibinfo  {journal} {Phys. Rev. Lett.}\ }\textbf {\bibinfo
  {volume} {82}},\ \bibinfo {pages} {4536} (\bibinfo {year}
  {1999})}\BibitemShut {NoStop}%
\bibitem [{\citenamefont {Hagiwara}\ \emph {et~al.}(2005)\citenamefont
  {Hagiwara}, \citenamefont {Regnault}, \citenamefont {Zheludev}, \citenamefont
  {Stunault}, \citenamefont {Metoki}, \citenamefont {Suzuki}, \citenamefont
  {Suga}, \citenamefont {Kakurai}, \citenamefont {Koike}, \citenamefont
  {Vorderwisch},\ and\ \citenamefont {Chung}}]{HagiwaraPRL05}%
  \BibitemOpen
  \bibfield  {author} {\bibinfo {author} {\bibfnamefont {M.}~\bibnamefont
  {Hagiwara}}, \bibinfo {author} {\bibfnamefont {L.~P.}\ \bibnamefont
  {Regnault}}, \bibinfo {author} {\bibfnamefont {A.}~\bibnamefont {Zheludev}},
  \bibinfo {author} {\bibfnamefont {A.}~\bibnamefont {Stunault}}, \bibinfo
  {author} {\bibfnamefont {N.}~\bibnamefont {Metoki}}, \bibinfo {author}
  {\bibfnamefont {T.}~\bibnamefont {Suzuki}}, \bibinfo {author} {\bibfnamefont
  {S.}~\bibnamefont {Suga}}, \bibinfo {author} {\bibfnamefont {K.}~\bibnamefont
  {Kakurai}}, \bibinfo {author} {\bibfnamefont {Y.}~\bibnamefont {Koike}},
  \bibinfo {author} {\bibfnamefont {P.}~\bibnamefont {Vorderwisch}}, \ and\
  \bibinfo {author} {\bibfnamefont {J.-H.}\ \bibnamefont {Chung}},\ }\href
  {\doibase 10.1103/PhysRevLett.94.177202} {\bibfield  {journal} {\bibinfo
  {journal} {Phys. Rev. Lett.}\ }\textbf {\bibinfo {volume} {94}},\ \bibinfo
  {pages} {177202} (\bibinfo {year} {2005})}\BibitemShut {NoStop}%
\bibitem [{\citenamefont {Veillette}\ \emph {et~al.}(2005)\citenamefont
  {Veillette}, \citenamefont {James},\ and\ \citenamefont
  {Essler}}]{VeillettePRB05}%
  \BibitemOpen
  \bibfield  {author} {\bibinfo {author} {\bibfnamefont {M.~Y.}\ \bibnamefont
  {Veillette}}, \bibinfo {author} {\bibfnamefont {A.~J.~A.}\ \bibnamefont
  {James}}, \ and\ \bibinfo {author} {\bibfnamefont {F.~H.~L.}\ \bibnamefont
  {Essler}},\ }\href {\doibase 10.1103/PhysRevB.72.134429} {\bibfield
  {journal} {\bibinfo  {journal} {Phys. Rev. B}\ }\textbf {\bibinfo {volume}
  {72}},\ \bibinfo {pages} {134429} (\bibinfo {year} {2005})}\BibitemShut
  {NoStop}%
\bibitem [{\citenamefont {Suzuki}\ and\ \citenamefont
  {Suga}(2005)}]{SuzukiPRB05}%
  \BibitemOpen
  \bibfield  {author} {\bibinfo {author} {\bibfnamefont {T.}~\bibnamefont
  {Suzuki}}\ and\ \bibinfo {author} {\bibfnamefont {S.-i.}\ \bibnamefont
  {Suga}},\ }\href {\doibase 10.1103/PhysRevB.72.014434} {\bibfield  {journal}
  {\bibinfo  {journal} {Phys. Rev. B}\ }\textbf {\bibinfo {volume} {72}},\
  \bibinfo {pages} {014434} (\bibinfo {year} {2005})}\BibitemShut {NoStop}%
\bibitem [{\citenamefont {Stone}\ \emph {et~al.}(2007)\citenamefont {Stone},
  \citenamefont {Zaliznyak}, \citenamefont {Hong}, \citenamefont {Broholm},\
  and\ \citenamefont {Reich}}]{StoneNature06}%
  \BibitemOpen
  \bibfield  {author} {\bibinfo {author} {\bibfnamefont {M.~B.}\ \bibnamefont
  {Stone}}, \bibinfo {author} {\bibfnamefont {I.~A.}\ \bibnamefont
  {Zaliznyak}}, \bibinfo {author} {\bibfnamefont {T.}~\bibnamefont {Hong}},
  \bibinfo {author} {\bibfnamefont {C.~L.}\ \bibnamefont {Broholm}}, \ and\
  \bibinfo {author} {\bibfnamefont {D.~H.}\ \bibnamefont {Reich}},\ }\href
  {\doibase 10.1038/nature04593} {\bibfield  {journal} {\bibinfo  {journal}
  {Nature}\ }\textbf {\bibinfo {volume} {440}},\ \bibinfo {pages} {187}
  (\bibinfo {year} {2007})}\BibitemShut {NoStop}%
\bibitem [{\citenamefont {Zhitomirsky}(2006)}]{ZhitomirskyPRB06}%
  \BibitemOpen
  \bibfield  {author} {\bibinfo {author} {\bibfnamefont {M.~E.}\ \bibnamefont
  {Zhitomirsky}},\ }\href {\doibase 10.1103/PhysRevB.73.100404} {\bibfield
  {journal} {\bibinfo  {journal} {Phys. Rev. B}\ }\textbf {\bibinfo {volume}
  {73}},\ \bibinfo {pages} {100404} (\bibinfo {year} {2006})}\BibitemShut
  {NoStop}%
\bibitem [{\citenamefont {Masuda}\ \emph {et~al.}(2006)\citenamefont {Masuda},
  \citenamefont {Zheludev}, \citenamefont {Manaka}, \citenamefont {Regnault},
  \citenamefont {Chung},\ and\ \citenamefont {Qiu}}]{MasudaPRL06}%
  \BibitemOpen
  \bibfield  {author} {\bibinfo {author} {\bibfnamefont {T.}~\bibnamefont
  {Masuda}}, \bibinfo {author} {\bibfnamefont {A.}~\bibnamefont {Zheludev}},
  \bibinfo {author} {\bibfnamefont {H.}~\bibnamefont {Manaka}}, \bibinfo
  {author} {\bibfnamefont {L.-P.}\ \bibnamefont {Regnault}}, \bibinfo {author}
  {\bibfnamefont {J.-H.}\ \bibnamefont {Chung}}, \ and\ \bibinfo {author}
  {\bibfnamefont {Y.}~\bibnamefont {Qiu}},\ }\href {\doibase
  10.1103/PhysRevLett.96.047210} {\bibfield  {journal} {\bibinfo  {journal}
  {Phys. Rev. Lett.}\ }\textbf {\bibinfo {volume} {96}},\ \bibinfo {pages}
  {047210} (\bibinfo {year} {2006})}\BibitemShut {NoStop}%
\bibitem [{\citenamefont {Kolezhuk}\ and\ \citenamefont
  {Sachdev}(2006)}]{KolezhukPRL06}%
  \BibitemOpen
  \bibfield  {author} {\bibinfo {author} {\bibfnamefont {A.}~\bibnamefont
  {Kolezhuk}}\ and\ \bibinfo {author} {\bibfnamefont {S.}~\bibnamefont
  {Sachdev}},\ }\href {\doibase 10.1103/PhysRevLett.96.087203} {\bibfield
  {journal} {\bibinfo  {journal} {Phys. Rev. Lett.}\ }\textbf {\bibinfo
  {volume} {96}},\ \bibinfo {pages} {087203} (\bibinfo {year}
  {2006})}\BibitemShut {NoStop}%
\bibitem [{\citenamefont {Bibikov}(2007)}]{BibikovPRB07}%
  \BibitemOpen
  \bibfield  {author} {\bibinfo {author} {\bibfnamefont {P.~N.}\ \bibnamefont
  {Bibikov}},\ }\href {\doibase 10.1103/PhysRevB.76.174431} {\bibfield
  {journal} {\bibinfo  {journal} {Phys. Rev. B}\ }\textbf {\bibinfo {volume}
  {76}},\ \bibinfo {pages} {174431} (\bibinfo {year} {2007})}\BibitemShut
  {NoStop}%
\bibitem [{\citenamefont {Sylju\aa{}sen}(2008)}]{SyljuasenPRB08}%
  \BibitemOpen
  \bibfield  {author} {\bibinfo {author} {\bibfnamefont {O.~F.}\ \bibnamefont
  {Sylju\aa{}sen}},\ }\href {\doibase 10.1103/PhysRevB.78.180413} {\bibfield
  {journal} {\bibinfo  {journal} {Phys. Rev. B}\ }\textbf {\bibinfo {volume}
  {78}},\ \bibinfo {pages} {180413} (\bibinfo {year} {2008})}\BibitemShut
  {NoStop}%
\bibitem [{\citenamefont {L\"uscher}\ and\ \citenamefont
  {L\"auchli}(2009)}]{LuscherPRB09}%
  \BibitemOpen
  \bibfield  {author} {\bibinfo {author} {\bibfnamefont {A.}~\bibnamefont
  {L\"uscher}}\ and\ \bibinfo {author} {\bibfnamefont {A.~M.}\ \bibnamefont
  {L\"auchli}},\ }\href {\doibase 10.1103/PhysRevB.79.195102} {\bibfield
  {journal} {\bibinfo  {journal} {Phys. Rev. B}\ }\textbf {\bibinfo {volume}
  {79}},\ \bibinfo {pages} {195102} (\bibinfo {year} {2009})}\BibitemShut
  {NoStop}%
\bibitem [{\citenamefont {Chernyshev}\ and\ \citenamefont
  {Zhitomirsky}(2009)}]{ChernyshevPRB09}%
  \BibitemOpen
  \bibfield  {author} {\bibinfo {author} {\bibfnamefont {A.~L.}\ \bibnamefont
  {Chernyshev}}\ and\ \bibinfo {author} {\bibfnamefont {M.~E.}\ \bibnamefont
  {Zhitomirsky}},\ }\href {\doibase 10.1103/PhysRevB.79.144416} {\bibfield
  {journal} {\bibinfo  {journal} {Phys. Rev. B}\ }\textbf {\bibinfo {volume}
  {79}},\ \bibinfo {pages} {144416} (\bibinfo {year} {2009})}\BibitemShut
  {NoStop}%
\bibitem [{\citenamefont {Masuda}\ \emph {et~al.}(2010)\citenamefont {Masuda},
  \citenamefont {Kitaoka}, \citenamefont {Takamizawa}, \citenamefont {Metoki},
  \citenamefont {Kaneko}, \citenamefont {Rule}, \citenamefont {Kiefer},
  \citenamefont {Manaka},\ and\ \citenamefont {Nojiri}}]{MasudaPRB10}%
  \BibitemOpen
  \bibfield  {author} {\bibinfo {author} {\bibfnamefont {T.}~\bibnamefont
  {Masuda}}, \bibinfo {author} {\bibfnamefont {S.}~\bibnamefont {Kitaoka}},
  \bibinfo {author} {\bibfnamefont {S.}~\bibnamefont {Takamizawa}}, \bibinfo
  {author} {\bibfnamefont {N.}~\bibnamefont {Metoki}}, \bibinfo {author}
  {\bibfnamefont {K.}~\bibnamefont {Kaneko}}, \bibinfo {author} {\bibfnamefont
  {K.~C.}\ \bibnamefont {Rule}}, \bibinfo {author} {\bibfnamefont
  {K.}~\bibnamefont {Kiefer}}, \bibinfo {author} {\bibfnamefont
  {H.}~\bibnamefont {Manaka}}, \ and\ \bibinfo {author} {\bibfnamefont
  {H.}~\bibnamefont {Nojiri}},\ }\href {\doibase 10.1103/PhysRevB.81.100402}
  {\bibfield  {journal} {\bibinfo  {journal} {Phys. Rev. B}\ }\textbf {\bibinfo
  {volume} {81}},\ \bibinfo {pages} {100402} (\bibinfo {year}
  {2010})}\BibitemShut {NoStop}%
\bibitem [{\citenamefont {Fischer}\ \emph {et~al.}(2010)\citenamefont
  {Fischer}, \citenamefont {Duffe},\ and\ \citenamefont
  {Uhrig}}]{FischerNewJPhys10}%
  \BibitemOpen
  \bibfield  {author} {\bibinfo {author} {\bibfnamefont {T.}~\bibnamefont
  {Fischer}}, \bibinfo {author} {\bibfnamefont {S.}~\bibnamefont {Duffe}}, \
  and\ \bibinfo {author} {\bibfnamefont {G.~S.}\ \bibnamefont {Uhrig}},\ }\href
  {http://stacks.iop.org/1367-2630/12/i=3/a=033048} {\bibfield  {journal}
  {\bibinfo  {journal} {New J. of Phys.}\ }\textbf {\bibinfo {volume} {12}},\
  \bibinfo {pages} {033048} (\bibinfo {year} {2010})}\BibitemShut {NoStop}%
\bibitem [{\citenamefont {Syromyatnikov}(2010)}]{SyromyatnikovPRB10}%
  \BibitemOpen
  \bibfield  {author} {\bibinfo {author} {\bibfnamefont {A.~V.}\ \bibnamefont
  {Syromyatnikov}},\ }\href {\doibase 10.1103/PhysRevB.82.024432} {\bibfield
  {journal} {\bibinfo  {journal} {Phys. Rev. B}\ }\textbf {\bibinfo {volume}
  {82}},\ \bibinfo {pages} {024432} (\bibinfo {year} {2010})}\BibitemShut
  {NoStop}%
\bibitem [{\citenamefont {Stephanovich}\ and\ \citenamefont
  {Zhitomirsky}(2011)}]{StephanovichEPL11}%
  \BibitemOpen
  \bibfield  {author} {\bibinfo {author} {\bibfnamefont {V.~A.}\ \bibnamefont
  {Stephanovich}}\ and\ \bibinfo {author} {\bibfnamefont {M.~E.}\ \bibnamefont
  {Zhitomirsky}},\ }\href {http://stacks.iop.org/0295-5075/95/i=1/a=17007}
  {\bibfield  {journal} {\bibinfo  {journal} {EPL (Europhysics Letters)}\
  }\textbf {\bibinfo {volume} {95}},\ \bibinfo {pages} {17007} (\bibinfo {year}
  {2011})}\BibitemShut {NoStop}%
\bibitem [{\citenamefont {Fischer}\ \emph {et~al.}(2011)\citenamefont
  {Fischer}, \citenamefont {Duffe},\ and\ \citenamefont
  {Uhrig}}]{FischerEPL11}%
  \BibitemOpen
  \bibfield  {author} {\bibinfo {author} {\bibfnamefont {T.}~\bibnamefont
  {Fischer}}, \bibinfo {author} {\bibfnamefont {S.}~\bibnamefont {Duffe}}, \
  and\ \bibinfo {author} {\bibfnamefont {G.~S.}\ \bibnamefont {Uhrig}},\ }\href
  {http://stacks.iop.org/0295-5075/96/i=4/a=47001} {\bibfield  {journal}
  {\bibinfo  {journal} {EPL (Europhysics Letters)}\ }\textbf {\bibinfo {volume}
  {96}},\ \bibinfo {pages} {47001} (\bibinfo {year} {2011})}\BibitemShut
  {NoStop}%
\bibitem [{\citenamefont {Fischer}(2011)}]{FischerDiss}%
  \BibitemOpen
  \bibfield  {author} {\bibinfo {author} {\bibfnamefont {T.}~\bibnamefont
  {Fischer}},\ }\emph {\bibinfo {title} {Description of quasiparticle decay by
  continuous unitary transformations}},\ \href
  {http://t1.physik.tu-dortmund.de/uhrig/phd/phd_Fischer_Tim_2011.pdf} {Ph.D.
  thesis},\ \bibinfo  {school} {TU-Dortmund} (\bibinfo {year}
  {2011})\BibitemShut {NoStop}%
\bibitem [{\citenamefont {Doretto}\ and\ \citenamefont
  {Vojta}(2012)}]{DorettoPRB12}%
  \BibitemOpen
  \bibfield  {author} {\bibinfo {author} {\bibfnamefont {R.~L.}\ \bibnamefont
  {Doretto}}\ and\ \bibinfo {author} {\bibfnamefont {M.}~\bibnamefont
  {Vojta}},\ }\href {\doibase 10.1103/PhysRevB.85.104416} {\bibfield  {journal}
  {\bibinfo  {journal} {Phys. Rev. B}\ }\textbf {\bibinfo {volume} {85}},\
  \bibinfo {pages} {104416} (\bibinfo {year} {2012})}\BibitemShut {NoStop}%
\bibitem [{\citenamefont {Fuhrman}\ \emph {et~al.}(2012)\citenamefont
  {Fuhrman}, \citenamefont {Mourigal}, \citenamefont {Zhitomirsky},\ and\
  \citenamefont {Chernyshev}}]{FuhrmanPRB12}%
  \BibitemOpen
  \bibfield  {author} {\bibinfo {author} {\bibfnamefont {W.~T.}\ \bibnamefont
  {Fuhrman}}, \bibinfo {author} {\bibfnamefont {M.}~\bibnamefont {Mourigal}},
  \bibinfo {author} {\bibfnamefont {M.~E.}\ \bibnamefont {Zhitomirsky}}, \ and\
  \bibinfo {author} {\bibfnamefont {A.~L.}\ \bibnamefont {Chernyshev}},\ }\href
  {\doibase 10.1103/PhysRevB.85.184405} {\bibfield  {journal} {\bibinfo
  {journal} {Phys. Rev. B}\ }\textbf {\bibinfo {volume} {85}},\ \bibinfo
  {pages} {184405} (\bibinfo {year} {2012})}\BibitemShut {NoStop}%
\bibitem [{\citenamefont {Zhitomirsky}\ and\ \citenamefont
  {Chernyshev}(2013)}]{ZhitomirskyRMP13}%
  \BibitemOpen
  \bibfield  {author} {\bibinfo {author} {\bibfnamefont {M.~E.}\ \bibnamefont
  {Zhitomirsky}}\ and\ \bibinfo {author} {\bibfnamefont {A.~L.}\ \bibnamefont
  {Chernyshev}},\ }\href {\doibase 10.1103/RevModPhys.85.219} {\bibfield
  {journal} {\bibinfo  {journal} {Rev. Mod. Phys.}\ }\textbf {\bibinfo {volume}
  {85}},\ \bibinfo {pages} {219} (\bibinfo {year} {2013})}\BibitemShut
  {NoStop}%
\bibitem [{\citenamefont {Oh}\ \emph {et~al.}(2013)\citenamefont {Oh},
  \citenamefont {Le}, \citenamefont {Jeong}, \citenamefont {Lee}, \citenamefont
  {Woo}, \citenamefont {Song}, \citenamefont {Perring}, \citenamefont {Buyers},
  \citenamefont {Cheong},\ and\ \citenamefont {Park}}]{JoosungPRL13}%
  \BibitemOpen
  \bibfield  {author} {\bibinfo {author} {\bibfnamefont {J.}~\bibnamefont
  {Oh}}, \bibinfo {author} {\bibfnamefont {M.~D.}\ \bibnamefont {Le}}, \bibinfo
  {author} {\bibfnamefont {J.}~\bibnamefont {Jeong}}, \bibinfo {author}
  {\bibfnamefont {J.-H.}\ \bibnamefont {Lee}}, \bibinfo {author} {\bibfnamefont
  {H.}~\bibnamefont {Woo}}, \bibinfo {author} {\bibfnamefont {W.-Y.}\
  \bibnamefont {Song}}, \bibinfo {author} {\bibfnamefont {T.~G.}\ \bibnamefont
  {Perring}}, \bibinfo {author} {\bibfnamefont {W.~J.~L.}\ \bibnamefont
  {Buyers}}, \bibinfo {author} {\bibfnamefont {S.-W.}\ \bibnamefont {Cheong}},
  \ and\ \bibinfo {author} {\bibfnamefont {J.-G.}\ \bibnamefont {Park}},\
  }\href {\doibase 10.1103/PhysRevLett.111.257202} {\bibfield  {journal}
  {\bibinfo  {journal} {Phys. Rev. Lett.}\ }\textbf {\bibinfo {volume} {111}},\
  \bibinfo {pages} {257202} (\bibinfo {year} {2013})}\BibitemShut {NoStop}%
\bibitem [{\citenamefont {Mourigal}\ \emph {et~al.}(2013)\citenamefont
  {Mourigal}, \citenamefont {Fuhrman}, \citenamefont {Chernyshev},\ and\
  \citenamefont {Zhitomirsky}}]{MourigalPRB13}%
  \BibitemOpen
  \bibfield  {author} {\bibinfo {author} {\bibfnamefont {M.}~\bibnamefont
  {Mourigal}}, \bibinfo {author} {\bibfnamefont {W.~T.}\ \bibnamefont
  {Fuhrman}}, \bibinfo {author} {\bibfnamefont {A.~L.}\ \bibnamefont
  {Chernyshev}}, \ and\ \bibinfo {author} {\bibfnamefont {M.~E.}\ \bibnamefont
  {Zhitomirsky}},\ }\href {\doibase 10.1103/PhysRevB.88.094407} {\bibfield
  {journal} {\bibinfo  {journal} {Phys. Rev. B}\ }\textbf {\bibinfo {volume}
  {88}},\ \bibinfo {pages} {094407} (\bibinfo {year} {2013})}\BibitemShut
  {NoStop}%
\bibitem [{\citenamefont {Pfeuty}(1970)}]{PfeutyAnnPhys70}%
  \BibitemOpen
  \bibfield  {author} {\bibinfo {author} {\bibfnamefont {P.}~\bibnamefont
  {Pfeuty}},\ }\href
  {http://www.sciencedirect.com/science/article/pii/0003491670902708}
  {\bibfield  {journal} {\bibinfo  {journal} {Annals of Physics}\ }\textbf
  {\bibinfo {volume} {57}},\ \bibinfo {pages} {79 } (\bibinfo {year}
  {1970})}\BibitemShut {NoStop}%
\bibitem [{\citenamefont {Chakrabati}\ \emph {et~al.}(1996)\citenamefont
  {Chakrabati}, \citenamefont {Dutta},\ and\ \citenamefont {Sen}}]{TFIMBook}%
  \BibitemOpen
  \bibfield  {author} {\bibinfo {author} {\bibfnamefont {B.~K.}\ \bibnamefont
  {Chakrabati}}, \bibinfo {author} {\bibfnamefont {A.}~\bibnamefont {Dutta}}, \
  and\ \bibinfo {author} {\bibfnamefont {P.}~\bibnamefont {Sen}},\ }\href@noop
  {} {\emph {\bibinfo {title} {Quantum Ising Phases and Transitions in
  Transverse Ising Models}}}\ (\bibinfo  {publisher} {Springer},\ \bibinfo
  {year} {1996})\BibitemShut {NoStop}%
\bibitem [{\citenamefont {Sachdev}(1999)}]{SachdevBook}%
  \BibitemOpen
  \bibfield  {author} {\bibinfo {author} {\bibfnamefont {S.}~\bibnamefont
  {Sachdev}},\ }\href@noop {} {\emph {\bibinfo {title} {Quantum Phase
  Transitions}}}\ (\bibinfo  {publisher} {Cambridge University Press},\
  \bibinfo {year} {1999})\BibitemShut {NoStop}%
\bibitem [{\citenamefont {Coldea}\ \emph {et~al.}(2010)\citenamefont {Coldea},
  \citenamefont {Tennant}, \citenamefont {Wheeler}, \citenamefont {Wawrzynska},
  \citenamefont {Prabhakaran}, \citenamefont {Telling}, \citenamefont
  {Habicht}, \citenamefont {Smeibidl},\ and\ \citenamefont
  {Kiefer}}]{ColdeaScience10}%
  \BibitemOpen
  \bibfield  {author} {\bibinfo {author} {\bibfnamefont {R.}~\bibnamefont
  {Coldea}}, \bibinfo {author} {\bibfnamefont {D.~A.}\ \bibnamefont {Tennant}},
  \bibinfo {author} {\bibfnamefont {E.~M.}\ \bibnamefont {Wheeler}}, \bibinfo
  {author} {\bibfnamefont {E.}~\bibnamefont {Wawrzynska}}, \bibinfo {author}
  {\bibfnamefont {D.}~\bibnamefont {Prabhakaran}}, \bibinfo {author}
  {\bibfnamefont {M.}~\bibnamefont {Telling}}, \bibinfo {author} {\bibfnamefont
  {K.}~\bibnamefont {Habicht}}, \bibinfo {author} {\bibfnamefont
  {P.}~\bibnamefont {Smeibidl}}, \ and\ \bibinfo {author} {\bibfnamefont
  {K.}~\bibnamefont {Kiefer}},\ }\href {\doibase 10.1126/science.1180085}
  {\bibfield  {journal} {\bibinfo  {journal} {Science}\ }\textbf {\bibinfo
  {volume} {327}},\ \bibinfo {pages} {177} (\bibinfo {year}
  {2010})}\BibitemShut {NoStop}%
\bibitem [{\citenamefont {Zamolodchikov}(1989)}]{ZamolodchikovIntJModPhysA89}%
  \BibitemOpen
  \bibfield  {author} {\bibinfo {author} {\bibfnamefont {A.~B.}\ \bibnamefont
  {Zamolodchikov}},\ }\href {\doibase 10.1142/S0217751X8900176X} {\bibfield
  {journal} {\bibinfo  {journal} {Int. J. Mod. Phys. A}\ }\textbf {\bibinfo
  {volume} {04}},\ \bibinfo {pages} {4235} (\bibinfo {year}
  {1989})}\BibitemShut {NoStop}%
\bibitem [{\citenamefont {{Morris}}\ \emph {et~al.}(2014)\citenamefont
  {{Morris}}, \citenamefont {{Vald{\'e}s Aguilar}}, \citenamefont {{Ghosh}},
  \citenamefont {{Koohpayeh}}, \citenamefont {{Krizan}}, \citenamefont
  {{Cava}}, \citenamefont {{Tchernyshyov}}, \citenamefont {{McQueen}},\ and\
  \citenamefont {{Armitage}}}]{MorrisPRL14}%
  \BibitemOpen
  \bibfield  {author} {\bibinfo {author} {\bibfnamefont {C.~M.}\ \bibnamefont
  {{Morris}}}, \bibinfo {author} {\bibfnamefont {R.}~\bibnamefont {{Vald{\'e}s
  Aguilar}}}, \bibinfo {author} {\bibfnamefont {A.}~\bibnamefont {{Ghosh}}},
  \bibinfo {author} {\bibfnamefont {S.~M.}\ \bibnamefont {{Koohpayeh}}},
  \bibinfo {author} {\bibfnamefont {J.}~\bibnamefont {{Krizan}}}, \bibinfo
  {author} {\bibfnamefont {R.~J.}\ \bibnamefont {{Cava}}}, \bibinfo {author}
  {\bibfnamefont {O.}~\bibnamefont {{Tchernyshyov}}}, \bibinfo {author}
  {\bibfnamefont {T.~M.}\ \bibnamefont {{McQueen}}}, \ and\ \bibinfo {author}
  {\bibfnamefont {N.~P.}\ \bibnamefont {{Armitage}}},\ }\href {\doibase
  10.1103/PhysRevLett.112.137403} {\bibfield  {journal} {\bibinfo  {journal}
  {Phys. Rev. Lett.}\ }\textbf {\bibinfo {volume} {112}},\ \bibinfo {pages}
  {137403} (\bibinfo {year} {2014})}\BibitemShut {NoStop}%
\bibitem [{\citenamefont {Kj\"all}\ \emph {et~al.}(2011)\citenamefont
  {Kj\"all}, \citenamefont {Pollmann},\ and\ \citenamefont
  {Moore}}]{KjallPRB11}%
  \BibitemOpen
  \bibfield  {author} {\bibinfo {author} {\bibfnamefont {J.~A.}\ \bibnamefont
  {Kj\"all}}, \bibinfo {author} {\bibfnamefont {F.}~\bibnamefont {Pollmann}}, \
  and\ \bibinfo {author} {\bibfnamefont {J.~E.}\ \bibnamefont {Moore}},\ }\href
  {\doibase 10.1103/PhysRevB.83.020407} {\bibfield  {journal} {\bibinfo
  {journal} {Phys. Rev. B}\ }\textbf {\bibinfo {volume} {83}},\ \bibinfo
  {pages} {020407} (\bibinfo {year} {2011})}\BibitemShut {NoStop}%
\bibitem [{\citenamefont {Essler}\ and\ \citenamefont
  {Konik}(2005)}]{EsslerReview05}%
  \BibitemOpen
  \bibfield  {author} {\bibinfo {author} {\bibfnamefont {F.~H.~L.}\
  \bibnamefont {Essler}}\ and\ \bibinfo {author} {\bibfnamefont {R.~M.}\
  \bibnamefont {Konik}},\ }in\ \href@noop {} {\emph {\bibinfo {booktitle} {From
  Fields to Strings: Circumnavigating Theoretical Physics}}},\ \bibinfo
  {editor} {edited by\ \bibinfo {editor} {\bibfnamefont {A.}~\bibnamefont
  {Vainshtein}}\ and\ \bibinfo {editor} {\bibfnamefont {J.}~\bibnamefont
  {Wheater}}}\ (\bibinfo  {publisher} {World Scientific},\ \bibinfo {address}
  {Singapore},\ \bibinfo {year} {2005})\ \Eprint
  {http://arxiv.org/abs/cond-mat/0412421} {arXiv:cond-mat/0412421} \BibitemShut
  {NoStop}%
\bibitem [{\citenamefont {Hamer}\ \emph {et~al.}(2006)\citenamefont {Hamer},
  \citenamefont {Oitmaa}, \citenamefont {Weihong},\ and\ \citenamefont
  {McKenzie}}]{HamerPRB06}%
  \BibitemOpen
  \bibfield  {author} {\bibinfo {author} {\bibfnamefont {C.~J.}\ \bibnamefont
  {Hamer}}, \bibinfo {author} {\bibfnamefont {J.}~\bibnamefont {Oitmaa}},
  \bibinfo {author} {\bibfnamefont {Z.}~\bibnamefont {Weihong}}, \ and\
  \bibinfo {author} {\bibfnamefont {R.~H.}\ \bibnamefont {McKenzie}},\ }\href
  {\doibase 10.1103/PhysRevB.74.060402} {\bibfield  {journal} {\bibinfo
  {journal} {Phys. Rev. B}\ }\textbf {\bibinfo {volume} {74}},\ \bibinfo
  {pages} {060402} (\bibinfo {year} {2006})}\BibitemShut {NoStop}%
\bibitem [{\citenamefont {James}\ \emph {et~al.}(2009)\citenamefont {James},
  \citenamefont {Goetze},\ and\ \citenamefont {Essler}}]{JamesPRB09}%
  \BibitemOpen
  \bibfield  {author} {\bibinfo {author} {\bibfnamefont {A.~J.~A.}\
  \bibnamefont {James}}, \bibinfo {author} {\bibfnamefont {W.~D.}\ \bibnamefont
  {Goetze}}, \ and\ \bibinfo {author} {\bibfnamefont {F.~H.~L.}\ \bibnamefont
  {Essler}},\ }\href {\doibase 10.1103/PhysRevB.79.214408} {\bibfield
  {journal} {\bibinfo  {journal} {Phys. Rev. B}\ }\textbf {\bibinfo {volume}
  {79}},\ \bibinfo {pages} {214408} (\bibinfo {year} {2009})}\BibitemShut
  {NoStop}%
\bibitem [{\citenamefont {Zaliznyak}\ and\ \citenamefont
  {Tranquada}(2013{\natexlab{a}})}]{ZaliznyakBook}%
  \BibitemOpen
  \bibfield  {author} {\bibinfo {author} {\bibfnamefont {I.}~\bibnamefont
  {Zaliznyak}}\ and\ \bibinfo {author} {\bibfnamefont {J.~M.}\ \bibnamefont
  {Tranquada}},\ }\href@noop {} {\emph {\bibinfo {title} {Neutron Scattering
  and Its Application to Strongly Correlated Systems}}},\ edited by\ \bibinfo
  {editor} {\bibfnamefont {A.}~\bibnamefont {Avella}}\ and\ \bibinfo {editor}
  {\bibfnamefont {F.}~\bibnamefont {Mancini}}\ (\bibinfo  {publisher}
  {Springer},\ \bibinfo {year} {2013})\BibitemShut {NoStop}%
\bibitem [{\citenamefont {Zaliznyak}\ and\ \citenamefont
  {Tranquada}(2013{\natexlab{b}})}]{ZaliznyakArxiv13}%
  \BibitemOpen
  \bibfield  {author} {\bibinfo {author} {\bibfnamefont {I.}~\bibnamefont
  {Zaliznyak}}\ and\ \bibinfo {author} {\bibfnamefont {J.~M.}\ \bibnamefont
  {Tranquada}},\ }\href@noop {} {\bibfield  {journal} {\bibinfo  {journal}
  {ArXiv e-prints}\ } (\bibinfo {year} {2013}{\natexlab{b}})},\ \Eprint
  {http://arxiv.org/abs/1304.4214} {arXiv:1304.4214} \BibitemShut {NoStop}%
\bibitem [{\citenamefont {Press}\ \emph {et~al.}(2005)\citenamefont {Press},
  \citenamefont {Teukolsky}, \citenamefont {Vetterling},\ and\ \citenamefont
  {Flannery}}]{NR}%
  \BibitemOpen
  \bibfield  {author} {\bibinfo {author} {\bibfnamefont {W.~H.}\ \bibnamefont
  {Press}}, \bibinfo {author} {\bibfnamefont {S.~A.}\ \bibnamefont
  {Teukolsky}}, \bibinfo {author} {\bibfnamefont {W.~T.}\ \bibnamefont
  {Vetterling}}, \ and\ \bibinfo {author} {\bibfnamefont {B.~P.}\ \bibnamefont
  {Flannery}},\ }\href@noop {} {\emph {\bibinfo {title} {Numerical Recipes in
  C++}}}\ (\bibinfo  {publisher} {Cambridge University Press},\ \bibinfo {year}
  {2005})\BibitemShut {NoStop}%
\bibitem [{\citenamefont {Gagliano}\ and\ \citenamefont
  {Balseiro}(1987)}]{GaglianoPRL87}%
  \BibitemOpen
  \bibfield  {author} {\bibinfo {author} {\bibfnamefont {E.~R.}\ \bibnamefont
  {Gagliano}}\ and\ \bibinfo {author} {\bibfnamefont {C.~A.}\ \bibnamefont
  {Balseiro}},\ }\href {\doibase 10.1103/PhysRevLett.59.2999} {\bibfield
  {journal} {\bibinfo  {journal} {Phys. Rev. Lett.}\ }\textbf {\bibinfo
  {volume} {59}},\ \bibinfo {pages} {2999} (\bibinfo {year}
  {1987})}\BibitemShut {NoStop}%
\bibitem [{\citenamefont {Dagotto}(1994)}]{DagottoRMP94}%
  \BibitemOpen
  \bibfield  {author} {\bibinfo {author} {\bibfnamefont {E.}~\bibnamefont
  {Dagotto}},\ }\href {\doibase 10.1103/RevModPhys.66.763} {\bibfield
  {journal} {\bibinfo  {journal} {Rev. Mod. Phys.}\ }\textbf {\bibinfo {volume}
  {66}},\ \bibinfo {pages} {763} (\bibinfo {year} {1994})}\BibitemShut
  {NoStop}%
\bibitem [{\citenamefont {Peierls}(1991)}]{PeierlsMoreSuprises}%
  \BibitemOpen
  \bibfield  {author} {\bibinfo {author} {\bibfnamefont {R.}~\bibnamefont
  {Peierls}},\ }\href@noop {} {\emph {\bibinfo {title} {More Surprises in
  Theoretical Physics}}}\ (\bibinfo  {publisher} {Princeton University Press},\
  \bibinfo {year} {1991})\BibitemShut {NoStop}%
\bibitem [{Note1()}]{Note1}%
  \BibitemOpen
  \bibinfo {note} {One may think that off-diagonal elements of the $g$-tensor
  might have the same effect. However, as a result of the local symmetry point
  group at the Co$^{2+}$ site (two-fold rotation axis around $b$), the $b$-axis
  is a principle axis of the $g$-tensor so an external magnetic field applied
  strictly along the $b$-axis does not induce a longitudinal field
  component.}\BibitemShut {Stop}%
\bibitem [{\citenamefont {Kunimoto}\ \emph {et~al.}(1999)\citenamefont
  {Kunimoto}, \citenamefont {Nagasaka}, \citenamefont {Nojiri}, \citenamefont
  {Luther}, \citenamefont {Motokawa}, \citenamefont {Ohta}, \citenamefont
  {Goto}, \citenamefont {Okubo},\ and\ \citenamefont {Kohn}}]{KunimotoJPSJ99}%
  \BibitemOpen
  \bibfield  {author} {\bibinfo {author} {\bibfnamefont {T.}~\bibnamefont
  {Kunimoto}}, \bibinfo {author} {\bibfnamefont {K.}~\bibnamefont {Nagasaka}},
  \bibinfo {author} {\bibfnamefont {H.}~\bibnamefont {Nojiri}}, \bibinfo
  {author} {\bibfnamefont {S.}~\bibnamefont {Luther}}, \bibinfo {author}
  {\bibfnamefont {M.}~\bibnamefont {Motokawa}}, \bibinfo {author}
  {\bibfnamefont {H.}~\bibnamefont {Ohta}}, \bibinfo {author} {\bibfnamefont
  {T.}~\bibnamefont {Goto}}, \bibinfo {author} {\bibfnamefont {S.}~\bibnamefont
  {Okubo}}, \ and\ \bibinfo {author} {\bibfnamefont {K.}~\bibnamefont {Kohn}},\
  }\href {\doibase 10.1143/JPSJ.68.1703} {\bibfield  {journal} {\bibinfo
  {journal} {Journal of the Physical Society of Japan}\ }\textbf {\bibinfo
  {volume} {68}},\ \bibinfo {pages} {1703} (\bibinfo {year}
  {1999})}\BibitemShut {NoStop}%
\end{thebibliography}%

\end{document}